\documentclass[preprint,showpacs,preprintnumbers,amsmath,amssymb]{revtex4}%

\usepackage{graphicx}
\usepackage{dcolumn}
\usepackage{bm}
\usepackage{epsfig}
\usepackage{amsmath}
\usepackage{amsfonts}
\usepackage{amssymb}%
\usepackage{caption, subcaption, floatrow}
\usepackage{float}
\usepackage{rotating}

\providecommand{\U}[1]{\protect\rule{.1in}{.1in}}

\newcommand{\be}{\begin{equation}}
\newcommand{\en}{\end{equation}}
\newcommand{\bea}{\begin{eqnarray}}
\newcommand{\ena}{\end{eqnarray}}

\newcommand{\lsim}{\lesssim}

\begin{document}
\title{Reconstructing k-inflation from $n_s(N)$ and reheating constraints}

\author{Ram\'on Herrera}
\email{ramon.herrera@pucv.cl}

\author{Michel Housset}
\email{michel.housset.b@mail.pucv.cl}

\author{Constanza Osses}
\email{constanza.osses.g@mail.pucv.cl}

\author{Nelson Videla}
\email{nelson.videla@pucv.cl} 

\affiliation{ Instituto de
F\'{\i}sica, Pontificia Universidad Cat\'{o}lica de Valpara\'{\i}so,
Avenida Brasil 2950, Casilla 4059, Valpara\'{\i}so, Chile.}


\date{\today}
\begin{abstract}
Inspired by the reconstruction scheme of the inflaton field potential $V(\phi)$ from the attractors
$n_s(N)$, we investigate the viability of reconstruct the inflationary potential within the framework of k-inflation for a non-linear kinetic term $K(X)=k_{n+1}X^n$ through three expressions for
the scalar spectral index $n_s(N)$, namely: (i) $n_s-1=-\frac{2}{N}$, (ii) $n_s-1=-\frac{p}{N}$, and (iii) $n_s-1=-\frac{\beta}{N^q}$. For each reconstructed potential, we determine the
 values of the parameter space which characterize it by requiring that it must reproduce the
observable parameters from PLANCK 2018 and BICEP/Keck results. Furthermore, we analyze the reheating
era by assuming a constant equation of state, 
in which we derive the relations between the reheating duration, the temperature at the end of reheating together with the reheating epoch, and the number of $e$-folds during inflation. In this sense, we unify the inflationary observables in order to narrow the parameter space
of each model within the framework of the reconstruction  in k-inflation.

\end{abstract}


\maketitle

\section{Introduction}



Inflation \cite{starobinsky1,inflation1,inflation2,inflation3} is
the most successful framework to describe the physics of the very early universe. The reason
for this is twofold. First, the inflationary universe offers a natural explanation to several long-standing puzzles of the Hot Big-Bang (HBB) model, such as the horizon, flatness, monopole problems, etc. Besides, since quantum fluctuations during the inflationary era may give rise to the primordial density perturbations \cite{Starobinsky:1979ty,R2,R202,R203,R204,R205}, inflation provides us with a causal interpretation of the origin of the temperature anisotropies observed on the Cosmic Microwave Background (CMB) \cite{COBE:1992syq,WMAP:2003elm,Planck:2013pxb,WMAP:2012nax,Planck:2013jfk,planck1,planck2,bicep1,bicep2}, while at the same time it comes with a mechanism to explain the Large-Scale Structure (LSS) of the Universe \cite{Abazajian:2013vfg}. The simplest single-field slow-roll inflation model, based on a canonical kinetic term and a potential $V(\phi)$, predicts a spectrum of the primordial curvature perturbations which is almost Gaussian
and almost scale-invariant, being confirmed by CMB observations (see Refs.\cite{Lidsey:1995np,Lyth:1998xn,Bassett:2005xm,Baumann:2009ds,Martin:2013tda,Renaux-Petel:2015bja} for reviews). In general, inflationary observables as the scalar spectral index $n_s$ and the tensor-to-scalar ratio $r$ are sensitive to the shape of the inflaton potential, and they also are strictly constrained by the Planck data in combination with
other cosmological data \cite{bicep2}. As far as tensor-to-scalar ratio is concerned, currently there only exists an
upper bound on it, since the tensor perturbations signature on the CMB as B-mode polarization has not been measured yet. The recent BICEP/Keck results \cite{BICEP:2021xfz} put a stringent bound on $r$, given by $r_{0.05}<0.036$ at 95$\%$ C.L. These new results have relevant consequences on the development of inflation models, since e.g., the power-law chaotic $V(\phi)\sim \phi^p$ is completely ruled out for every $p$ in the canonical framework, as well as the standard version of natural inflation \cite{Freese:1990rb,Adams:1992bn}. Therefore, it would be challenging to build models which predict a tensor-to-scalar ratio at the current
detection limit. For a further discussion regarding the detection prospects of next generation CMB experiments, see Ref.\cite{Achucarro:2022qrl} (and references therein).

After inflation ends there is a phase of reheating \cite{reh1,reh2}, where the universe fills with standard model (SM) particles. These particles interact and will eventually thermalise to equilibrium at a reheating temperature $T_{re}$, and then the standard hot big-bang cosmology of radiation dominated era follows. From the theoretical point of view, the reheating temperature is assumed to be larger than the temperature of Electroweak (EW) transition \cite{Bassett:2005xm,reh3,reh4}. For some conservative issues, the temperature for the reheating era must be much larger by several orders of magnitude than the temperature reached in the big-bang nucleosynthesis (BBN), i.e., above a MeV \cite{bound1,bound2}. Historically, reheating was first studied by means 
perturbative decays of the inflaton field \cite{reh1,reh2}. However, the transition from inflation to the standard hot big-bang cosmology could happen via very different mechanisms than the perturbative reheating approach. In particular, parametric resonance effects may be significant under certain
regimes, particularly early in the oscillating regime when the oscillation amplitude is
large \cite{Greene:1997fu,reh4}. As it was shown in Ref. \cite{Podolsky:2005bw}, the out-of-equilibrium nonlinear dynamics of fields produces a sharp variation of the equation-of-state (EoS) parameter during the reheating phase. Consequently, the physics of reheating is complicated, highly uncertain, and in addition it cannot be directly probed by observations. Nevertheless, one may obtain indirect constraints on reheating 
by assuming for the fluid a constant equation of state (EoS) $w_{re}$ during reheating. In this sense,
then we find certain relations between the reheating temperature $T_{re}$ and the duration of reheating $N_{re}$ with $w_{re}$ and inflationary observables \cite{Martin:2010kz,paper1,Martin:2014nya,paper2,Cai:2015soa,paper3,Rehagen:2015zma,Ueno:2016dim,Panotopoulos:2020qzi,Mishra:2021wkm,Osses:2021snt,Gong:2015qha}
On theoretical and observational grounds, going beyond standard canonical inflation
within General Relativity (GR) has become of a special interest. A more general scenario is provided by the k-inflation framework, in which a non-linear function of the kinetic term $K(X)$ is present in the Lagrangian \cite{kinflation,ps}, i.e. $P(\phi,X)=K(X)-V(\phi)$, where $X=-\frac{1}{2}\partial_{\mu}\phi\,\partial^{\mu}\phi$ is the canonical kinetic term. A non-trivial kinetic term can lead a reduced scalar propagation speed ($c_s^2<1$). This introduces new features, including a suppression of the tensor-to-scalar ratio $r$ \cite{ps} and at the same time a large amount of non-Gaussianities (NG) \cite{Chen:2006nt,DeFelice:2011uc,DeFelice:2011zh}. As a particular case, $K(X)=k_{n+1}X^n$ \cite{Mukhanov:2005bu,liddle,Unnikrishnan:2012zu}, with $n$ taking integer values and $k_{n+1}$ being constants such that $k_{n+1}X^n$ has units of $M_{pl}^4$, accounts for a reduced scalar propagation speed $c_s^2=\dfrac{1}{2n-1}$ if $n>1$. Thus, at least in principle, k-inflation becomes phenomenologically distinguishable from standard inflation, where $c_s^2=1$. In this regard, in Refs.\cite{Mishra:2022ijb,Pareek:2021lxz}, the authors found that the simplest power-law  potentials, as the chaotic quadratic and quartic ones, become compatible with current bounds on the tensor-to-scalar ratio in a non-canonical scenario with $K(X)\propto X^n$ if $n>1$. Besides, scalar field models with a non-trivial kinetic term have been discussed in the context of k-essence in order to explain the observed speed-up of the universe at present time \cite{Chiba:1999ka,Armendariz-Picon:2000nqq,Armendariz-Picon:2000ulo,Deffayet:2011gz}.

As usual, inflation is studied on the basis of a potential for which, within the slow-roll approximation, inflationary observables as the spectral index $n_s$ and the tensor-to-scalar ratio $r$ are computed and then compared with CMB observations through the $n_s-r$ plane. However, in Ref.\cite{chiba} it was proposed to reconstruct the potential by assuming a standard canonical inflaton field from an attractor for the spectral index as a function of the number of $e$-folds $N$, i.e., $n_s(N)$ in the framework of Einstein gravity. This inverse problem is motivated by the observational data regarding the spectral index, since the attractor $n_s-1=-\,2/N$, which is predicted from Starobinsky model \cite{starobinsky}, the $\alpha$-attractor models \cite{alpha1,alpha2,alpha3}, the quadratic chaotic inflation model \cite{inflation3}, and Higgs inflation with a non-minimal coupling \cite{higgs1,higgs2} is compatible with latest data of PLANCK. Similar reconstruction procedures have been also proposed 
previously. In \cite{Mukhanov:2013tua}, the inflaton potential $V(\phi)$ is reconstructed from a certain function of the slow-roll parameter $\epsilon (N)$, while in \cite{Roest:2013fha} the potential $V(N)$ is obtained when both slow-roll parameters
$\epsilon (N)$ and $\eta (N)$ are given, and then the tensor-to-scalar ratio is computed. Also, more general expressions $n_s-1=-\,p/N$ and $n_s-1=-\,\beta/N^q$ have been studied in Refs.\cite{Garcia-Bellido:2014gna,pattractor} and \cite{Huang:2007qz}, respectively. This reconstruction scheme have been also applied in
theories beyond the standard framework, e.g. in k-inflation, when a coupling to the standard kinetic term $K(\phi)X$ is considered \cite{Yi:2021xhw,Herrera:2020mjh}, Randall-Sundrum II braneworld \cite{Herrera:2019xhs,Bhattacharya:2019ryo}, G-inflation \cite{Herrera:2018mvo}, and warm inflation \cite{Herrera:2018cgi}.

The main goal of the present work is to apply the reconstruction scheme from the scalar spectral index $n_s(N)$ in the framework of k-inflation, from a non-linear kinetic term $K(X)=k_{n+1}X^n$, in order to obtain the potential of the inflaton field $V(\phi)$. In fact, we study how k-inflation modifies
the reconstruction  of the inflationary scenario by assuming as attractor the scalar spectral index $n_s(N)$.
Furthermore, we discuss the implications for reheating phase regarding its duration and temperature.

We organize our work as follows: After this introduction,
in the next section we briefly presents the dynamics of inflation and the 
the reconstruction scheme from the scalar spectral index $n_s(N)$ within the framework
of k-inflation, as well as the basic
formulas for determining the duration of reheating as well as
for the reheating temperature after inflation. In Sections \ref{attractor1}, \ref{attractor2}, and \ref{attractor3} we analyze the expressions $n_s-1=-2/N$, $n_s-1=-\,p/N$, and $n_s-1=-\beta/N^q$, respectively. For all examples associated to $n_s(N)$, we use the approximation of perfect fluid with a constant equation of state in order to study reheating after inflation. Finally, in Section \ref{conclusions} we summarize our 
findings and present the conclusions. We choose units of $c=\hbar=1$.

\section{k-inflation: Theoretical Framework}\label{kinflation}

\subsection{Dynamics of a generalized scalar field}

Our framework is four-dimensional GR
coupled to a single scalar field with a generalized matter Lagrangian
$\mathcal{L}(\phi,X)$, where $\phi$ is the scalar field and 
$X=-\frac{1}{2}\partial_{\mu}\phi\,\partial^{\mu}\phi$ is the standard kinetic term. is described by the action
\begin{equation}\label{action}
    \mathcal{S}= \int d^4x \: \sqrt{-g} \left( \frac{M_{pl}^2}{2}R+\mathcal{L}(\phi,X)\right),
\end{equation}
where $g$ is the determinant of metric tensor $g_{\mu\nu}$, $M_{pl}=2.4 \times 10^{18}~GeV$ is the reduced Planck mass, and $R$ is the Ricci scalar. For this matter Lagrangian, the energy-momentum tensor can be recast into the form of
the energy-momentum tensor for a perfect fluid. Accordingly, the pressure $P$ and energy density $\rho$ are given in terms of $\phi$ and $X$ as follows \cite{ps}
\begin{eqnarray}
P(\phi,X)&=&\mathcal{L},\label{pressure}\\
\rho(\phi,X)&=&2X \mathcal{L}_{,X}-\mathcal{L},\label{density}
\end{eqnarray}
respectively. For a canonical scalar field, we have $P(\phi,X)=X-V(\phi)$, where $V(\phi)$ is the 
effective potential associated to the scalar field. In
this work, we restrict ourselves on the case in which the Lagrangian of the scalar field is of
the form
\begin{equation}\label{kessence}
    P(\phi,X)=K(X)-V(\phi),
\end{equation}
with $K(X)$ being an arbitrary function of the kinetic term. For this lagrangian, the corresponding energy density is given by 
\begin{equation}\label{density1}
    \rho(\phi,X)=2X K_{,X}-K+V.
\end{equation}
In the following, the notation $P_{,X}$ corresponds to
$\frac{\partial P}{\partial X}$, $K_{,X}=\frac{\partial K}{\partial X}$, $V_{,\phi}=\frac{\partial V}{\partial \phi}$, $P_{,XX}=\frac{\partial ^2 P}{\partial X^2}$, etc.

The dynamics of our system is described by the Friedmann equation and the equation for energy conservation
\begin{eqnarray}
H^2&=&\dfrac{\rho}{3M_{pl}^2},\label{H2}\\
\dot{\rho}&=&-3H(\rho+P),\label{cont}
\end{eqnarray}
where, in order to write Eq.(\ref{H2}) we have assumed a spatially flat Friedmann-Lema\^{i}tre-Robertson-Walker (FLRW) metric.

Defining the sound speed $c_s$, which describes the properties of the scalar field in the fluid description, we have
\begin{equation}\label{cs}
    c_s^2\equiv \dfrac{P_{,X}}{\rho_{,X}}=\dfrac{P_{,X}}{2XP_{,XX}+P_{,X}}.
\end{equation}
Also, we can write the energy density conservation as
\begin{equation}
    \ddot{\phi}+3c_s^2 H\dot{\phi}+\frac{\rho_{,\phi}}{\rho_{,X}}=0,\label{KGk}
\end{equation}
which is a generalized version of the usual Klein-Gordon in a FLRW background. Here we have considered
a homogeneous scalar field i.e., $\phi(\vec{r},t)=\phi(t)$.

\subsection{Inflationary dynamics}

In this subsection, we will analyze the slow-roll approximation in order to describe the inflationary 
universe.

For k-inflation, the slow-roll conditions read as follows \cite{Mukhanov:2005bu}
\begin{eqnarray}
XK_{,X}\ll V,\,\,K\ll V,\,\,\big |\ddot{\phi}\big |\ll\frac{V_{,\phi}}{\rho_{,X}},\label{srcond}
\end{eqnarray}
which ensure an inflationary expansion driven by the potential. Following Ref.\cite{liddle}, during the slow-phase, the inflationary dynamics
can be quantified in terms of the three parameters
\begin{eqnarray}
    \epsilon &=&-\dfrac{\dot{H}}{H^2}\label{epsilon},\\ \eta &=&-\dfrac{\dot{\epsilon}}{H \epsilon},\label{eta}\\
    s &=&\dfrac{\dot{c_s}}{H c_s}.\label{s}
\end{eqnarray}
It is assumed that during slow-roll inflation the parameter $\left(\epsilon,\big |\eta\big |,\big |s\big |\right)\ll1$.
Therefore, by considering $P(\phi,X)=K(X)-V(\phi)$, the Friedmann equation (\ref{H2}) and field equation for the scalar field (\ref{cont}) reduce to
\begin{eqnarray}
3M_{pl}^2H^2&\simeq & V(\phi),\label{H2sr}\\
3c_s^2 H\dot{\phi}+\frac{V_{,\phi}}{K_{,X}}&\simeq &0,\label{contsr}
\end{eqnarray}
since we neglect the acceleration term $\ddot{\phi}$
in the equation of motion for the scalar ﬁeld and $V$ is the dominant term in the expression for the energy density.

For our concrete non-canonical scalar field model without specifying the potential, in the following we will assume that the Lagrangian density and energy density are given by
\begin{eqnarray}
  P(\phi,X)&=&k_{n+1}X^n-V(\phi),\label{kessence2}\\
  \rho(\phi,X)&=&(2n-1)k_{n+1}X^n+V(\phi).\label{kessence3}
\end{eqnarray}
Here, the power $n$ takes integer real values and $k_{n+1}$ is a coupling constant such that $k_{n+1}X^n$ has units of $M_{pl}^4$. In particular, when $n=1$ and $k_2=1$, we recover the standard Lagrangian density for a canonical scalar field. By replacing Eq.\eqref{kessence2} into \eqref{cs}, we find that the sound speed is a constant and given by
\begin{equation}
    c_s^2=\dfrac{1}{2n-1}.
\end{equation}
It should be noted that it can be very small for large $n$, in which $n\gg 1$, and for the specific case when $n=1$ (canonical field), we have $c_s^2=1$.

Under the slow-roll approximation, the field equation for the scalar field $\phi$ from Eq.(\ref{contsr}) takes the form
\begin{equation}\label{continuity2}
    \dot{\phi}V_{,\phi}\simeq -6nHk_{n+1}X^n.
\end{equation}

From Eqs.(\ref{epsilon}) and (\ref{eta}), we can introduce the slow-roll parameters $\epsilon_V$ and 
$\eta_V$ in terms of the potential and its derivatives with respect to the scalar field as \cite{liddle}
\begin{eqnarray}
    &\epsilon_V\simeq -\dfrac{1}{2}\dfrac{V_{,\phi}}{V}\dfrac{\dot{\phi}}{H},\label{epsilonv}\\
    &\eta_V \simeq-\dfrac{V_{,\phi \phi}}{V_{,\phi}}\dfrac{\dot{\phi}}{H}.\label{etav}
\end{eqnarray}

Now, by combining $X=\dfrac{\dot{\phi}^2}{2}$ and Eqs.(\ref{H2}) and (\ref{continuity2}), we find that the ratio $\frac{\dot{\phi}}{H}$ becomes
\begin{equation}\label{ratio}
    \dfrac{\dot{\phi}}{H}=-\alpha(n)\left(\dfrac{V_{,\phi}}{V^n}\right)^{\frac{1}{2n-1}},
\end{equation}
where the quantity $\alpha(n)$ is defined as
\begin{equation}\label{alpha}
    \alpha(n)=\left(\dfrac{6^{n-1}M_{pl}^{2n}}{n\,k_{n+1}}\right)^{\frac{1}{2n-1}}.
\end{equation}
Then, the slow-roll parameters can be written as
\begin{eqnarray}
    \epsilon_V&=&\dfrac{\alpha(n)}{2}\left(\dfrac{V_{,\phi}^{2n}}{V^{3n-1}}\right)^{\frac{1}{2n-1}}\label{epsilonalpha},\\
    \eta_V&=&\alpha(n)\left(\dfrac{V_{,\phi \phi}^{2n-1}}{V_{,\phi}^{2n-2}V^n}\right)^{\frac{1}{2n-1}}\label{etaalpha}.
\end{eqnarray}
Notice that for this model the slow-roll parameter $s$ becomes zero, since the sound speed is a constant. Also, we have that for the case $n=1$ we recover the usual slow-roll parameters associated to a canonical scalar field.

The amount of inflation is expressed in terms of the number of $e$-folds $N$, defined by
\begin{equation}\label{N}
    N\equiv\ln\left(\dfrac{a(t_{end})}{a(t)}\right)=\int_{t}^{t_{end}}H\,dt=\int_{\phi}^{\phi_{end}}H\,\dfrac{d\phi}{\dot{\phi}}\simeq \dfrac{1}{\alpha (n)}\int_{\phi_{end}}^{\phi}\left(\frac{V^n}{V_{,\phi}}\right)^{\frac{1}{2n-1}}\,d\phi,
\end{equation}
where $t$ and $t_{end}$ correspond to two different values of cosmic time; $t_{end}$ denotes the end of inflation, and
$t$ corresponds to the time when the cosmological scales cross the Hubble-radius. In order to write $N$ as an integral over the scalar field, we have used Eq.(\ref{ratio}) and we have also assumed that the number
of $e$-folds at the end of inflation is $N_{end}=0$.

\subsection{Perturbations}

In the following we shall brieﬂy review cosmological perturbations in the model of k-inflation.

Under the slow-roll approximation, the primordial scalar power spectrum was derived in \cite{ps}, and it results
\begin{equation}
    \mathcal{P}_S=\dfrac{1}{8\pi^2M_{pl}^2c_s}\dfrac{H^2}{\epsilon_V},\label{Psv}
\end{equation}
which is evaluated at the time of horizon exit at $c_sk=aH$ ($k$ is a comoving wavenumber). Besides, the scalar spectral index $n_s$ is defined as
\begin{equation}
n_s-1= \left.\frac{d \ln \mathcal{P}_S}{d \ln k}\right|_{c_sk=aH}\simeq-2\epsilon-\eta-s.\label{nsk}
\end{equation}
On the other hand, the tensor power spectrum $\mathcal{P}_T$ and the corresponding spectral index $n_T$
are given by \cite{ps}
\begin{equation}
\mathcal{P}_T=\frac{2H^2}{\pi^2M_{pl}^2},\quad \quad n_T=
\left.\frac{d \ln \mathcal{P}_T}{d \ln k}\right|_{c_sk=aH}.\label{Ptk}
\end{equation}
From Eqs.(\ref{Psv}) and (\ref{Ptk}), the tensor-to-scalar ratio is found to be
\begin{equation}
r\equiv \frac{\mathcal{P}_T}{\mathcal{P}_S}=16 \epsilon c_s=-8 c_s n_T.\label{rt}
\end{equation}
As it can be seen from Eq.(\ref{rt}), the consistency relation is modified in comparison to standard inflation 
($c_s=1$). Thus, at least in principle, k-inflation is phenomenologically distinguishable from standard inflation \cite{ps}.

By replacing Eqs.(\ref{H2sr}) and (\ref{epsilonalpha}) in (\ref{Psv}), the scalar power spectrum for our particular Lagrangian density (\ref{kessence2}) becomes
\begin{equation}\label{Ps}
    \mathcal{P}_{\mathcal{S}}\simeq\dfrac{1}{12\pi^2M_{pl}^4c_s\alpha(n)}\left(\dfrac{V^{5n-2}}{V_{,\phi}^{\,2n}}\right)^{\frac{1}{2n-1}},
\end{equation}
and its spectral index can be expressed as a function of slow-roll parameters given by Eqs.\eqref{epsilonv} and \eqref{etav} as follows
\begin{equation}\label{nsVN}
    n_s-1\simeq\dfrac{1}{2n-1}\left[2n\eta_V-2(5n-2)\epsilon_V\right].
\end{equation}
For a given potential $V(\phi)$, we can determine the
values of the parameters characterizing the model by requiring that it must reproduce the
observable values of $\mathcal{P}_S$, $n_s$ and the upper bound on $r$. However, we are not going to concern
ourselves with a particular choice of $V(\phi)$, since in the present work our
main interest is to reconstruct the inflationary potential from a given $n_s(N)$ within the framework of k-inflation. 

\subsection{Reconstructing $V(\phi)$ in k-inflation}

Let us to explain how to reconstruct $V(\phi)$ from $n_s(N)$ within k-inflation for a non-linear kinetic term $K(X)=k_{n+1}X^n$, following Refs.\cite{liddle} and \cite{chiba}.

From Eq.\eqref{N} an important relation arises between the number of $e$-folds, the potential and its derivatives with respect to the scalar field, namely
\begin{equation}\label{Nvphi}
   dN=\dfrac{1}{\alpha (n)}\left(\frac{V^n}{V_{,\phi}}\right)^{\frac{1}{2n-1}}\,d\phi.
\end{equation}
In order to obtain become real quantities, we impose that $V_{,\phi}>0$. Therefore, we can use Eq.\eqref{Nvphi} to obtain a relation between the derivative of the potential with respect to the number of $e$-folds $N$ and the scalar field such that
\begin{equation}\label{vphivn}
    V_{,N}=\alpha (n)\left(\dfrac{V_{,\phi}^2}{V}\right)^{\frac{n}{2n-1}}.
\end{equation}
Note that here we have that $V_{,N}>0$. In order to find the relation between the scalar field $\phi$ and $N$, we combine Eqs.\eqref{Nvphi} and \eqref{vphivn} such that
\begin{equation}
    \phi(N)=\alpha(n)^{\frac{2n-1}{2n}}\int \left(\dfrac{V_{,N}}{V^n}\right)^{\frac{1}{2n}}\,dN\label{phiN}.
\end{equation}
Now, by replacing Eqs.\eqref{Nvphi} and \eqref{vphivn} in \eqref{epsilonv} and \eqref{etav}, we rewrite the slow-roll parameters $\epsilon_V$ and $\eta_V$ in terms of the derivatives of the potential with respect to the number of $e$-folds as
\begin{eqnarray}
 \epsilon_V&=&\dfrac{1}{2}\dfrac{V_{,N}}{V}\label{epsilonn},\\
 \eta_V&=&\left(\dfrac{2n-1}{2n}\right)\dfrac{V_{,NN}}{V_{,N}}+\dfrac{V_{,N}}{2V}\label{etan},
\end{eqnarray}
respectively. Then, the scalar spectral index \eqref{nsVN}
is rewritten as
\begin{eqnarray}
  n_s-1&=&-\dfrac{2V_{,N}}{V}+\dfrac{V_{,NN}}{V_{,N}}=\ln\left(\dfrac{V_{,N}}{V^2}\right)_{,N}\label{ns}.
\end{eqnarray}

We also may write the tensor-to-scalar ratio as
\begin{equation}
r=16 \epsilon_V c_s=\frac{8}{\sqrt{2n-1}}\frac{V_{,N}}{V}.\label{rN}
\end{equation}

In such a way, Eqs.\eqref{phiN} and \eqref{ns} 
are the main equations which enables to us to reconstruct 
$V(\phi)$ from the attractor $n_s(N)$.

\subsection{Reheating}

Here we shall briefly describe how to derive the expressions for the number of
$e$-folds during the reheating epoch $N_{re}$ as well as the reheating temperature $T_{re}$ for k-inflation 
considering the reconstruction scheme. For the derivation of the main formulas, we mainly follow
Refs.\cite{paper1,paper2,paper3} and also \cite{Pareek:2021lxz}.
First, it is assumed that during reheating phase the dominant contribution to the energy density of the Universe comes from a component which has an  effective equation-of-state parameter (EoS) $w_{re}$, and its energy density can be related to the scale factor through $\rho\propto a^{-3(1+w_{re})}$. Here, we have considered that the EoS parameter $w_{re}$ is constant. Therefore, we can write down the following relation
\begin{equation}
    \frac{\rho_{end}}{\rho_{re}}=\left(\frac{a_{end}}{a_{re}}\right)^{-3(1+w_{re})},\label{rhoendre}
\end{equation}
where the subscript ``$end$\," denotes the end of inflation and ``$re$" the end of reheating phase. Besides, the number of $e$-folds of reheating $N_{re}$ may be related to the scale factors at the end of inflation and reheating according to
\begin{equation}
    e^{-N_{re}}=\frac{a_{end}}{a_{re}}\label{enre}.
\end{equation}

Then, by combining Eqs. \eqref{rhoendre} and \eqref{enre}, we can write the number
of $e$-folds during the reheating scenario as 
\begin{equation}
    N_{re}=\frac{1}{3(1+w_{re})}\ln\left(\frac{\rho_{end}}{\rho_{re}}\right), \,\,\textup{with}\,\,w_{re}\neq -1.\label{nrerho}
\end{equation}

On the other hand, by combining the time derivative of Friedmann Eq.\eqref{H2} and the conservation equation
for energy density \eqref{cont} along with the slow-roll parameter $\epsilon$ \eqref{epsilon}, and then replacing the energy density and pressure from Eqs. \eqref{kessence2} and \eqref{kessence3}, the slow-roll parameter $\epsilon$ becomes
\begin{equation}
    \epsilon=\dfrac{3nk_{n+1}X^n}{(2n-1)k_{n+1}X^n+V}.
\end{equation}
If we solve the latter equation for $\dfrac{X^n}{V}$, we have
\begin{equation}\label{xv}
    \dfrac{X^n}{V}=\dfrac{\epsilon}{k_{n+1}(3n-(2n-1)\epsilon)}.
\end{equation}
In addition, Eq. \eqref{kessence3} can be rewritten as
\begin{equation}\label{density3}
    \rho=V\left[(2n-1)k_{n+1}\dfrac{X^n}{V}+1\right].
\end{equation}
In such a way, we can express the energy density as a function of the slow-roll parameter $\epsilon$ and the scalar field potential $V$ by replacing Eq. \eqref{xv} into \eqref{density3}
\begin{equation}
    \rho=V\left[\dfrac{(2n-1)\epsilon}{3n-(2n-1)\epsilon}+1\right].
\end{equation}
Accordingly, the relationship between the energy density and the potential at
the end of inflation ($\epsilon(\phi_{end})=1$) is given by
\begin{equation}\label{rhoendV}
    \rho_{end}=\frac{3n}{n+1}V_{end},
\end{equation}
where $V(\phi=\phi_{end})=V_{end}$. Note that this result for the energy density $\rho_{end}$ becomes 
different from those already used in \cite{Pareek:2021lxz}, where $\rho_{end}=\frac{3}{2}V_{end}$, which corresponds to standard canonical inflation ($n=1$) \cite{paper1,paper2,paper3}.

After replacement of (\ref{rhoendV}) in Eq. \eqref{nrerho}, we get that $N_{re}$ becomes
\begin{equation}
N_{re}=\frac{1}{3(1+w_{re})}\ln\left(\frac{3n}{n+1}\frac{V_{end}}{\rho_{re}}\right).\label{nrein}
\end{equation}
At the end of reheating phase, the energy density of the universe is assumed to 
have the form of a relativistic fluid
\begin{equation}
\rho_{re}=\frac{\pi^2}{30}g_{re}T_{re}^4,\label{rho_re}
\end{equation}
where $g_{re}$ is the number of internal degrees of freedom of relativistic particles
at the end of reheating. Assuming that the degrees of freedom come from Standard Model (SM) particles, then $g_{re}\sim {\mathcal{O}}(100)$ for a temperature $T \gtrsim 175$\,GeV \cite{Husdal:2016haj}, while for a Minimal Supersymmetric Standard Model (MSSM), we have that $g_{re}\sim {\mathcal{O}}(200)$ \cite{Adhikari:2019uaw}.

Regarding the entropy, its density is defined as 
\begin{equation}
    s=\frac{2\pi^2}{45}gT^3 ,\label{s}
\end{equation}
where the temperature is inversely proportional to the scale factor for radiation epoch, i.e., $T\propto a^{-1}$. Then, by replacing the last relation in Eq.\eqref{s}, we have that $s\propto a^{-3}$. If we assume the conservation
of entropy, i.e., $gT^3a^3=const.$, then by applying this conservation between reheating and the present time, results
\begin{equation}
    g_{re}T_{re}^3a_{re}^3=g_0T_0^3 a_0^3,\label{re0}
\end{equation}
where $g_0$ denotes the number of internal degrees of freedom of relativistic particles
today, which comes from photons and neutrinos and $T_0=2.725$ K is the temperature of the
universe today. Then, Eq.(\ref{re0}) becomes \cite{paper1}
\begin{equation}
    g_{re}T_{re}^3=\left(\frac{a_0}{a_{re}}\right)^3\left[2T_0^3 + \frac{21}{4}T_{\nu 0}^3\right].\label{conservation_s}
\end{equation}
Here, $T_{\nu 0}$ is the neutrino temperature at present time. For the contribution coming from neutrinos at the right-hand side of (\ref{conservation_s}), we have used that $T_{\nu 0}=\left(\frac{4}{11}\right)^{1/3} T_0$. Besides, the ratio $\frac{a_0}{a_{re}}$ can be rewritten as 
\begin{equation}
\frac{a_0}{a_{re}}=\frac{a_0}{a_{eq}}\frac{a_{eq}}{a_{re}},
\end{equation}
where $\frac{a_{re}}{a_{eq}}$ can be related to the duration in $e$-folds of the radiation dominated epoch $N_{RD}$ according to $e^{-N_{RD}}=\frac{a_{re}}{a_{eq}}$. In this way, Eq.(\ref{conservation_s}) is rewritten
as
\begin{equation}
    T_{re}=T_0\left(\frac{a_0}{a_{eq}}\right)e^{N_{RD}}\left(\frac{43}{11g_{re}}\right).\label{T}
\end{equation}
For the ratio $\frac{a_0}{a_{eq}}$ we have (see, e.g. \cite{Pareek:2021lxz}) 
\begin{equation}
\frac{a_0}{a_{eq}}=\frac{a_0}{a_k}\frac{a_k}{a_{end}}\frac{a_{re}}{a_{eq}}\frac{a_{end}}{a_{re}}=\frac{a_0 H_k}{c_s k}e^{-N}e^{-N_{re}}e^{-N_{RD}},
\end{equation}
where we have considered that $\frac{a_k}{a_{end}}=e^{-N}$.
Also, in order to obtain the last equation, we have considered as well that the condition for horizon crossing is defined as $c_s\,k=a_kH_k$.
Using this result in \eqref{T} we find
\begin{equation}
    T_{re}=\left(\frac{43}{11g_{re}}\right)^{1/3}\left(\frac{a_0 T_0}{c_s k}\right) H_k e^{-N} e^{-N_{re}}.\label{trein}
\end{equation}
Assuming $g_{re}\sim \mathcal{O}(100)$ and the pivot scale $\frac{k}{a_0}=0.05 \textup{ Mpc}^{-1}$ by PLANCK, the number of $e$-folds during reheating is obtained after replacing of Eq. (\ref{trein}) in Eq. \eqref{nrein}, which results
\begin{equation}
N_{re}=\frac{4}{1-3\,w_{re}}\left[61.643 - \dfrac{1}{4}\ln\left(\dfrac{3n}{n+1}\right) - \ln \left(\frac{c_s\,V_{end}^{\,1/4}}{H_k}\right) - N\right],\label{nre}
\end{equation}
where $H_k$ can be written down using the expression for the tensor-to-scalar ratio $r=P_T/P_S$. Finally, combining Eqs.\eqref{nrein} and (\ref{rho_re}), the reheating temperature is computed as follows
\begin{equation}
T_{re}=\left(\frac{30}{g_{re}\,\pi^2}\right)^{1/4}\,\left(\frac{3n}{n+1}\,V_{end}\right)^{1/4}\,e^{-\frac{3}{4}(1+w_{re})N_{re}}.\label{tre}
\end{equation}
The model-dependent part of the main equations for reheating (\ref{nre}) and (\ref{tre}) are  the sound speed $c_s$, and therefore the power associated to the kinetic term $n$. Also, the potential at the end of inflation $V_{end}$, the Hubble rate when the cosmological scale crosses the Hubble radius $H_k$, and the number of $e$-folds $N_k$. Note that both $N_{re}$ and $T_{re}$ implicitly depend on the observables $\mathcal{P}_s$, $r$ and $n_s$. It is also remarkable that the canonical case is recovered when $n=1$.

In the following we will study different relations between the scalar spectral index $n_s$ in terms of the number of $e$-folds $N$ (attractors) in order to reconstruct the inflationary scenario.

\section{First attractor $n_s-1=-\dfrac{2}{N}$}\label{attractor1}
\subsection{Dynamics of inflation}
In order to obtain concrete results, as a first example we consider the attractor $n_s-1=-\dfrac{2}{N}$ \cite{chiba}. In this case, Eq. \eqref{ns} becomes
\begin{equation}
    -\dfrac{2}{N}=\ln\left(\dfrac{V_{,N}}{V^2}\right)_{,N},\label{ns1}
\end{equation}
which upon a first integration yields
\begin{equation}
    \dfrac{A}{N^2}=\dfrac{V_{,N}}{V^2}\label{alpha1},
\end{equation}
where $A$ is an integration constant, which is positive since $V_{,N}>0$. If we integrate once with respect to the number of $e$-folds, we find the following expression for the potential $V(N)$
\begin{equation}
    V(N)=\dfrac{N}{A+BN}\label{v1},
\end{equation}
with $B$ being a second integration constant, which can be either $B  \gtrless 0$ or $B=0$. In order to obtain an expression for the scalar field as a function of the number of $e$-folds $N$ for any value of the power $n$, we replace the potential \eqref{v1} into Eq. \eqref{phiN}, yielding
\begin{eqnarray}
    \phi(N)-C_1&=&\alpha ^{\frac{2n-1}{2n}}\int \left(\frac{A}{N^2}\right)^{\frac{1}{2n}}\left(\frac{A+BN}{N}\right)^{\frac{n-2}{2n}}dN\\
    &=& \mathcal{F}(N),\label{phi1}
\end{eqnarray}
where $C_1$ is a new integration constant and $\mathcal{F}(N)$ is a function
defined as
\begin{equation}
   \mathcal{F}(N) =\beta (N)\,{}_2F_1\left[\dfrac{1}{2},1,\dfrac{1}{n},-\dfrac{A}{BN}\right].\label{FN}
\end{equation}
Here ${}_2F_1$ corresponds to the hypergeometric function \cite{F21} and $\beta (N)$ is another function given by
\begin{equation}
    \beta (N)=\dfrac{n\,\alpha^{\footnotesize{\frac{2n-1}{2n}}}}{(n-1)B N}\left(\dfrac{A}{N^2}\right)^{\frac{1}{2n}}\left(\dfrac{A}{N}+B\right)^{\frac{3n-2}{2n}}.
\end{equation}
Accordingly, from Eq.(\ref{phi1}), the number of $e$-folds
expressed in terms of the scalar field can be written as
\begin{equation}
    N=\mathcal{F}^{-1}(\phi-C_1), \label{inversa}
\end{equation}
where $\mathcal{F}^{-1}$ represents the inverse function
of function (\ref{FN}).
In this way, by replacing Eq.(\ref{inversa}) into Eq.(\ref{v1}), the potential being a function of the scalar field is obtained as follows
\begin{equation}
   V(\phi)=\dfrac{\mathcal{F}^{-1}(\phi-C_1)}{A+B \mathcal{F}^{-1}(\phi-C_1)}\label{vphi} .
\end{equation}




As a particular case in which $n=1$, i.e., for standard inflation, the reconstructed potential (\ref{vphi}) becomes
\begin{equation}
    V(\phi)=\dfrac{1}{B}\tanh^2\left(\sqrt{\dfrac{B}{\alpha\,A}}\,\dfrac{(\phi-C_1)}{2}\right),
\end{equation}
and this potential corresponds to the T-model inflation studied in \cite{tmodel}. On the other hand, for the case $n=2$, the reconstructed potential (\ref{vphi}) reduces to
\begin{equation}
    V(\phi)=\dfrac{(\phi-C_1)^2}{4A\sqrt{\alpha^3 A}+B(\phi-C_1)^2}.\label{Vphi2}
\end{equation}
As asymptotic cases of Eq.(\ref{Vphi2}), first we observe that if $4A\sqrt{\alpha^3 A}\gg B(\phi-C_1)^2$, $V(\phi)$ behaves as a quadratic chaotic potential \cite{inflation3}
\begin{equation}
    V(\phi)\simeq \dfrac{(\phi-C_1)^2}{4A\sqrt{\alpha
    ^3 A}},
\end{equation}

while if $4A\sqrt{\alpha^3 A}\ll B(\phi-c)^2$, it becomes a constant potential
\begin{equation}
    V\simeq\dfrac{1}{B}.
\end{equation}

For values of the power $n$ such that $n>2$, Eq.(\ref{inversa}) cannot be solved analytically for $N=N(\phi)$, and therefore we are not able to obtain an analytical expression for $V(\phi)$.

\subsection{Cosmological perturbations}

If we replace Eq.(\ref{vphivn}) into Eq.\eqref{alpha1}, and using the expression for the scalar power spectrum \eqref{Ps}, we find that the first integration constant $A$ is constrained to be
\begin{equation}
    A=\dfrac{\sqrt{2n-1}N^2}{12\pi^2M_{pl}^4\mathcal{P}_s}.\label{AA}
\end{equation}
On the other hand, upon replacement of Eq.\eqref{v1} into Eq.\eqref{rN}, and considering the constraint on $A$ given by Eq.(\ref{AA}), we obtain that the tensor-to-scalar ratio may be expressed in terms of the number of $e$-folds
as follows
\begin{equation}
    r=\frac{8}{\sqrt{2n-1}N+12 B\pi^2 M_{pl}^4  \mathcal{P}_s}. \label{rN1}
\end{equation}
The allowed values of the integration constants $A$ and $B$ can be found by using the 
CMB constraints on the inflationary observables. In particular, we use
the constraint on the amplitude of the scalar spectrum $\mathcal{P}_s\simeq2.2\times 10^{-9}$ to set $A$, while the allowed values of $B$ are found from the upper bound on $r$ by the BICEP/Keck data, i.e., $r<0.036$ \cite{BICEP:2021xfz} for a range $N=50-70$. Therefore, if $r_0$ denotes the current upper limit in the tensor-to-scalar ratio, from Eq.(\ref{rN1}) we have that
 lower bound on $B$ yields
\begin{equation}
    B>\frac{8-\sqrt{2n-1}N r_0}{12 \pi^2 M_{pl}^4  \mathcal{P}_s r_0}.\label{Bvalues}
\end{equation}

In Table \ref{tab:AB1}, we summarize the different values associated to the integration constants $A$ and $B$ for the specific case in which the power $n$ associated to the kinetic term is $n=2$, from the observational parameters $\mathcal{P}_s$ and $r$. In this sense, we find that the constraints on the parameters $A$ and $B$ are $A\sim \mathcal{O}\left(10^{10}\right)$ and $B\gtrsim \mathcal{O}\left(10^{8}\right)$, respectively.

\begin{table}[H]
    \centering
    \captionsetup[subtable]{position = below}
    \captionsetup[table]{position=bottom}
    \begin{subtable}{0.5\textwidth}
        \begin{tabular}{|c|c|c|}
            \hline
            \hspace{5mm}$N$\hspace{5mm} & \hspace{5mm}$A\,[M_{pl}^{-4}]$\hspace{5mm} & \hspace{5mm}$B\,[M_{pl}^{-4}]$\hspace{5mm}  \\\hline
            50 & $1.66\times10^{10}$ & \,$B>5.20\times 10^8$\, \\
            60 & $2.39\times10^{10}$ & \,$B>4.54\times 10^8$\, \\
            70 & $3.26\times10^{10}$ & \,$B>3.88\times 10^8$\,\\\hline
        \end{tabular}
    
    \end{subtable}%
    \caption{Results for the constraints on the integration constants $A$ and $B$ for the first attractor $n_s-1=-\frac{2}{N}$ in  k-inflation scenario when $n=2$. Here we have used $\mathcal{P}_s\simeq2.2\times 10^{-9}$ and $r_0=0.036$}\label{tab:AB1}
\end{table}

\begin{figure}[H]
\centering
        \includegraphics[width=1\linewidth]{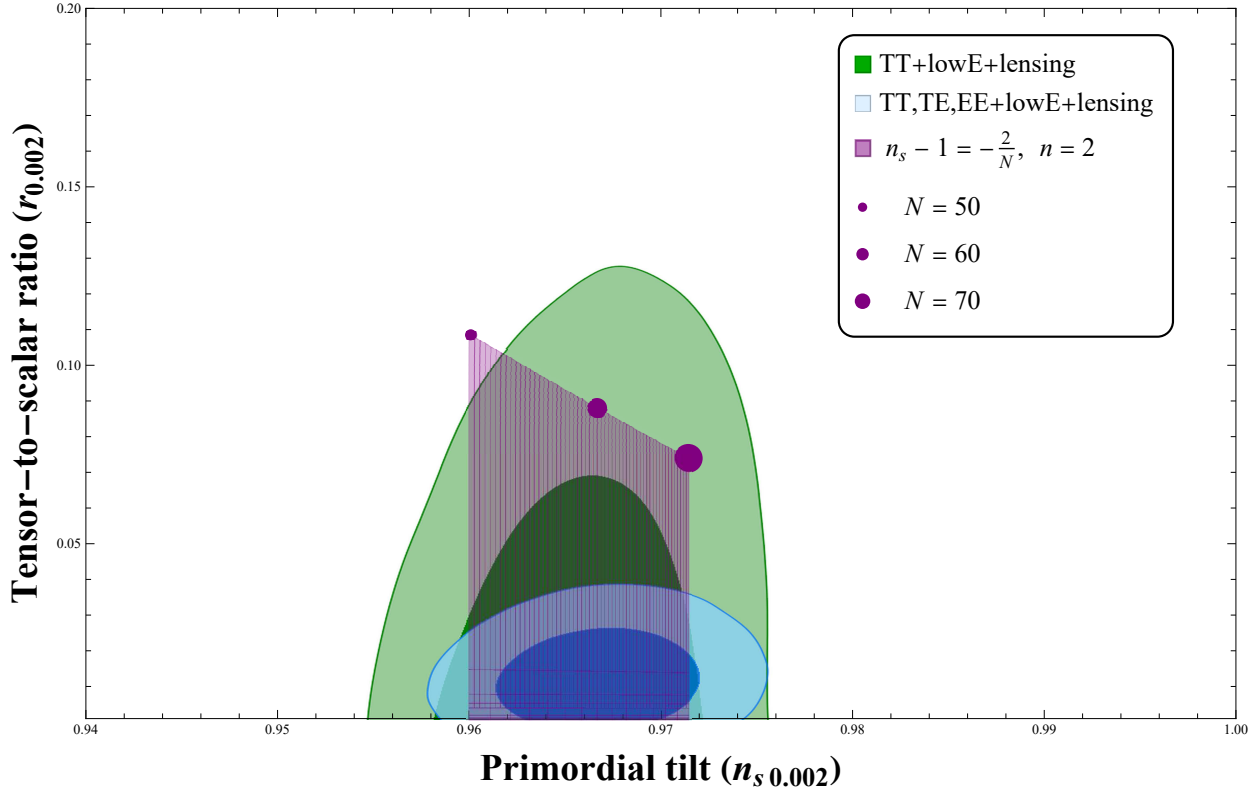}
    \caption{Plot of the tensor-to-scalar ratio $r$ against the scalar spectral index $n_s$ for the first attractor $n_s-1=-\frac{2}{N}$ using $n=2$ along with the two-dimensional marginalized joint confidence contours for $(n_s,r)$ at 68\% and 95\% C.L. from the latest BICEP/Keck results.}\label{rns1}
\end{figure}

Furthermore, the predictions on the $n_s-r$ plane may be generated by plotting  the attractor $n_s-1=-\dfrac{2}{N}$ and the tensor-to-scalar ratio given by Eq. \eqref{rN1}, by varying simultaneously the integration constant $B$ in a wide range and $N$ within the range $N=50-70$. Fig. \ref{rns1} shows the $n_s-r$ plane for $n=2$ considering the two-marginalized joint confidence at 68\% and 95\% C.L., from the latest BICEP/Keck results.

 In the next subsection, we will study the reheating constraints by means an effective EoS parameter $w_{re} $ together with the reconstruction of inflation (effective potential) in order to see whether the feasible parameter space of the model can be narrowed or not.

\subsection{Reheating}

\begin{figure}[H]
    \includegraphics[scale=0.32]{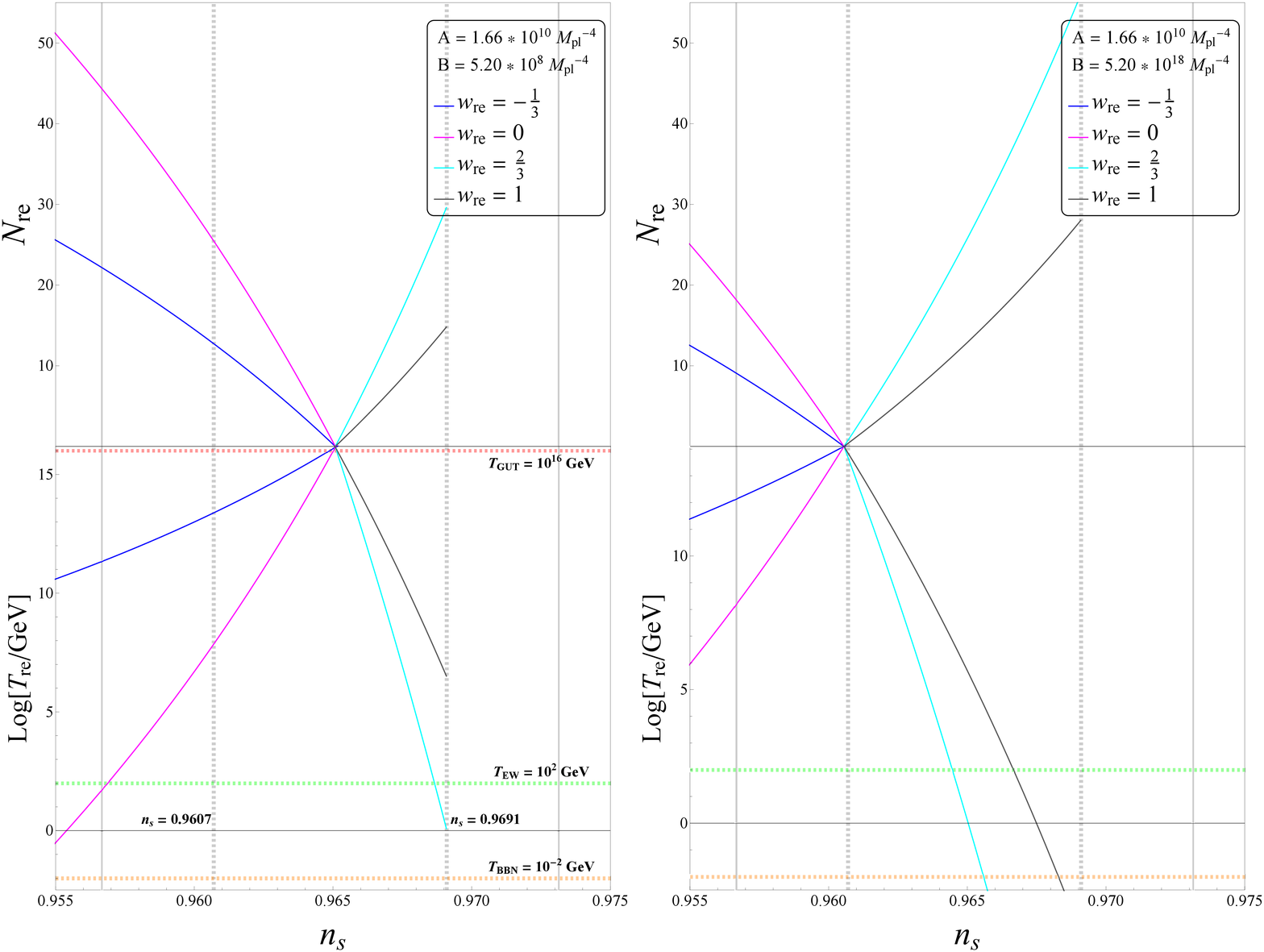}
    \caption{Plots of the reheating duration, $N_{re}$, and the reheating temperature $T_{re}$ for the power $n=2$ against the spectral index $n_s$ for the attractor $n_s-1=\frac{2}{N}$ for different values of the EoS parameter $w_{re}$. The blue, pink, cyan and black curves correspond to the EoS parameter $w_{re}=-1/3,0,2/3,1$, respectively. The gray dotted vertical lines indicates the observational constraints on the spectral index, i.e. $n_s=0.965\pm 0.004$. The green dotted horizontal line shows reheating temperatures being below the electroweak scale $T_{re}=10^{2}$ GeV, while the orange dotted line indicates the lower bound on $T_{re}$ from BBN, $T_{re}= 10^{-2}$ GeV. For the upper and lower left panels we have used $A=1.66\times10^{10}\,M_{pl}^{-4}$ and $B=5.20\times10^{8}\,M_{pl}^{-4}$, while for the upper and lower right panels we have used $A=1.66\times10^{10}\,M_{pl}^{-4}$ and $B=5.20\times10^{18}\,M_{pl}^{-4}$. }
    \label{pntre2}
\end{figure}

Now we are interested in studying the predictions of reheating phase regarding its duration through the number of $e$-folds $N_{re}$ as well as the temperature $T_{re}$. In doing so, we plot parametrically Eqs.\eqref{nre} and \eqref{tre} together with the expression for the attractor $n_s-1=-\frac{2}{N}$, with respect to the number of $e$-folds $N$ for several values of the EoS parameter $w_{re}$ over the range $-\frac{1}{3}\leq w_{re}\leq 1$. In Fig. \ref{pntre2}, we show the plots for the number of $e$-folds of the reheating phase $N_{re}$ (upper panel) and the reheating temperature $T_{re}$ (lower panel) against the scalar spectral index $n_s$ for
the case $n=2$ together with the allowed values of $A$ and the minimum value of $B$ at $N=50$ (see Table \ref{tab:AB1}). The behaviour of the reheating predictions become almost the same for the other values of the constants shown in Table \ref{tab:AB1}. However, we will restrict ourselves to show the plots that fit better with the current observational data. As a first finding, we mention that the maximum reheating temperature is given by $T_{re}\simeq 10^{16}$ GeV, which is similar to those found for a canonical scalar field, where $T_{re}\lsim 7\times 10^{15}$ GeV \cite{paper1}. Secondly, we observe that when $B$ increases and deviates from its minimum allowed value, the duration of reheating increases while the reheating temperature decreases from its maximum value. This 
behaviour can be inferred from the lower left and right panels of Fig.\ref{pntre2}, where the curves shift to the left and the maximum reheating temperature point decreases, respectively.

Assuming that reheating period can be parametrized by an effective EoS $P=w_{re}\rho$, it becomes important to distinguish what EoS parameter $w_{re}$ within the range $-\frac{1}{3}\leq w_{re}\leq 1$ is preferred by current observational bounds. Consequently, from the reheating temperature plots we observe the value of $n_s$ at which each curve enters the purple shaded region as well as the value when all curves converge. Thus, we will obtain the allowed ranges for both the scalar spectral index $n_s$  and the number of $e$-folds $N$ for fixed values of the integration constants $A$ and $B$. For completeness, the results found from the analysis of upper right and lower right plots of Fig. \ref{pntre2} are summarized in Table \ref{wn1}. We also mention that the values we have obtained for the number of $e$-folds 
during reheating phase reheating are similar to those found for $A$ and $B$ from Table \ref{tab:AB1}.

\begin{table}[H]
\centering
\begin{tabular}{| c | c |}
\hline
$w_{re}$ & $N$ \\ \hline
$-1/3$ & 51 - 57 \\
0 & 51 - 57\\
2/3 & 57 - 64\\
1 & 57 - 65 \\ \hline
\end{tabular} 
  \caption{Summary of the allowed range for the number of $e$-folds for each EoS parameter $w_{re}$ assuming the power $n$ associated to the kinetic term $n=2$.} \label{wn1}
\end{table}

\begin{figure}[H]
    \centering
    \includegraphics[scale=0.43]{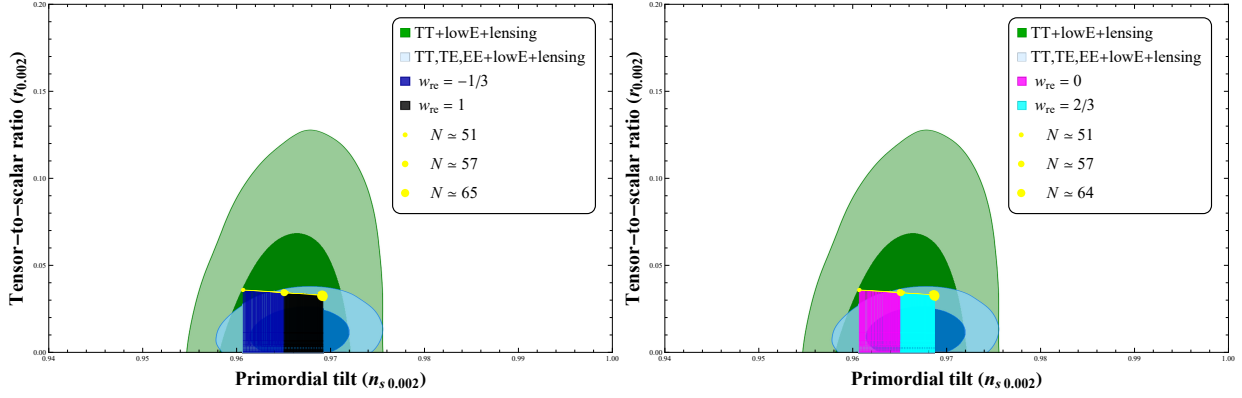}
    \caption{Predictions for the tensor-to-scalar ratio $r$ against the scalar spectral index $n_s$ for the first attractor $n_s-1=-\frac{2}{N}$ using $n=2$ after applying reheating constraints. The left panel corresponds to $w_{re}=-1/3$ and $w_{re}=1$, while the right panel for $w_{re}=0$ and $w_{re}=2/3$, respectively.}
    \label{fig:rns1new}
\end{figure}

As a consequence, if we restrict the number of $e$-folds $N$ (see Table \ref{wn1}) it is possible to obtain an allowed range for the reheating temperature for each parameter of state $w_{re}$: $10^{13}\textup{ GeV}\lsim T_{re} \lsim 10^{16}\textup{ GeV}$ for $w_{re}=-1/3$, $10^{7}\textup{ GeV}\lsim T_{re} \lsim 10^{16}\textup{ GeV}$ for $w_{re}=0$, $10^{2}\textup{ GeV}\lsim T_{re} \lsim 10^{16}\textup{ GeV}$ for $w_{re}=2/3$, and $10^{6}\textup{ GeV}\lsim T_{re} \lsim 10^{16}\textup{ GeV}$ for $w_{re}=1$. In order to obtain previous results, we have considered the lower bound of $B$ already found from the inflationary observables, see Table \ref{tab:AB1}. Once the number of $e$-folds $N$ becomes narrowed from reheating constraints, we are able to give the new predictions for this model on the $n_s-r$ plane. These new contour regions are displayed in Fig. \ref{fig:rns1new}. The left panel shows the predictions for both $w_{re}=-1/3$ (blue region) and $w_{re}=1$ (black region) while the right panel shows the predictions for $w_{re}=0$ (magenta region) and for $w_{re}=2/3$ (cyan region). It should be noted that for $w_{re}=-1/3$ and $w_{re}=0$ the predictions are the same, whereas for $w_{re}=2/3$ the region starts at the same point of the black region and it ends almost at the same point. In such a way, the four EoS parameters considered within the range $-\frac{1}{3}\leq w_{re}\leq 1$ are preferred by current observational bounds.

\section{Second attractor $n_s-1=-\dfrac{p}{N}$}\label{attractor2}
\subsection{Dynamics of inflation}

Now, we consider a little more general attractor $n_s-1=-\dfrac{p}{N}$ \cite{Garcia-Bellido:2014gna, pattractor}, where $p$ corresponds to a real number. By replacing in Eq. \eqref{ns}, we have that
\begin{equation}
    -\dfrac{p}{N}=\ln\left(\dfrac{V_{,N}}{V^2}\right)_{,N}\label{ns2}.
\end{equation}
A first integration of this equation with respect to the number of $e$-folds yields
\begin{equation}
    \dfrac{A}{N^p}=\dfrac{V_{,N}}{V^2}\label{AN},
\end{equation}
where $A$ is a positive integration constant. Integrating again with respect to $N$ we found that the potential as a function of $N$ can be written as
\begin{equation}
    V(N)=\dfrac{p-1}{AN^{1-p}+B(p-1)},\label{pneq1}
\end{equation}
where $B$ is a new integration constant and $p\neq 1$. 

As the case $p=1$ is a singular case, we will analyze it separately. Thus, for the case $p=1$, and in order to avoid that $V(N)=0$, we must first replace 
replace $p=1$ into Eq. \eqref{ns2} and then integrate with respect to the number of $e$-folds twice to find the potential
as a function of the number of $e$-folds $N$
 \begin{equation}\label{v2b}
    V(N)=\dfrac{1}{B-A\,\ln(N)}.
\end{equation}
We must note that this equation has a pole at $\ln N=\frac{B}{A}$ if $B$ is positive, so we can only consider the range $\ln\,N <\frac{B}{A}$ for that the potential to be positive. However, in the following we will assume $\ln N \ll \frac{B}{A}$ in order to obtain analytical solutions. On the other hand, analytically invertible expressions for $N(\phi)$ are only found for $n=2$. Therefore, from Eq.(\ref{phiN}) we can integrate to obtain $N(\phi)$ such that
\begin{equation}
    N(\phi)\simeq\left(\dfrac{3}{4}\right)^{4/3}\dfrac{(\phi-C_1)^{4/3}}{(\alpha^3 A)^{1/3}},
\end{equation}
where $C_1$ is a new integration constant such that $\phi > C_1$ in order to $N$ be real and positive. If we replace the latter equation into \eqref{v2b} and take the limit $\ln N\ll \frac{B}{A}$, the reconstructed potential $V(\phi)$ becomes
\begin{equation}\label{v2p1}
    V(\phi)\simeq \dfrac{1}{B}\left[1+\dfrac{4 A}{3 B}\ln\left(\frac{3\left (\phi-C_1\right)}{\left(\alpha ^3 A\right)^{1/4}}\right)\right],
\end{equation}
which is similar in form to the obtained in
loop inflation (LI) model, see Ref.\cite{Martin:2013tda}.

For the general case in which $p\neq 1$, we have that by replacing Eq.\eqref{v2} 
in (\ref{phiN}), the scalar field as a function of the number of $e$-folds becomes
\begin{eqnarray}
    \phi(N)-C_1&=&\alpha^{\frac{2n-1}{2n}}\int \left[\dfrac{A}{N^p}\left(\dfrac{p-1}{AN^{1-p}+B(p-1)}\right)^{2-n}\right]^{\frac{1}{2n}}\,dN\\
    &=& \mathcal{G}(N),\label{dphi2}
\end{eqnarray}
where $C_1$ is a new integration constant and $\mathcal{G}(N)$ is a function
defined as
\begin{equation}
   \mathcal{G}(N) = g(N)\,{}_2F_1\left[1,\dfrac{2-n(5-3p)-p}{2n(p-1)},1+\frac{2n-p}{2n(1-p)},-\dfrac{A N^{1-p}}{B (p-1)}\right].\label{FNp}
\end{equation}
Here ${}_2F_1$ corresponds to the hypergeometric function \cite{F21} and $g (N)$ is another function given by
\begin{equation}
    g(N)=\frac{\alpha (n)^{\frac{2n-1}{2n}}}{2n-p}2nN \left(A N^{-p}\right)^{\frac{1}{2n}}\left(\frac{p-1}{A N^{1-p}+B (p-1)}\right)^{\frac{2-3n}{2n}}.\label{gN}
\end{equation}
The number of $e$-folds as a function of the scalar field is obtained from Eq.(\ref{dphi2}) as follows
\begin{equation}
    N=\mathcal{G}^{-1}(\phi-C_1), \label{inversa2}
\end{equation}
where $\mathcal{G}^{-1}$ represents the inverse function
of function (\ref{gN}).
In this way, by replacing Eq.(\ref{inversa2}) into Eq.(\ref{pneq1}), the potential being a function of the scalar 
field for $p\neq 1$ becomes
\begin{equation}
    V(\phi)=\dfrac{p-1}{A \left(\mathcal{G}^{-1}(\phi-C_1)\right)^{1-p}+B(p-1)}\label{v2}.
\end{equation}
From Eqs.(\ref{inversa2}) and (\ref{v2}), we see that $N(\phi)$, and in turns $V(\phi)$, can not be expressed analytically for values of the power $n>2$ associated to the kinetic term. So, we will restrict ourselves to
the case $p\neq1$ for $n=2$.


Thus, from Eq.\eqref{inversa2} the number of $e$-folds $N(\phi)$ is found to be
\begin{equation}
    N(\phi)=\gamma (p) \left(\phi-C_1\right)^{\frac{4}{4-p}}, \,\,\textup{with}\,p\neq 4,\label{Nphip}
\end{equation}
where $\gamma (p)$ is defined as follows
\begin{equation}
    \gamma (p)=\left[\left(\frac{4-p}{4}\right)^{4}\frac{1}{\alpha ^3 A}\right]^{\frac{1}{4-p}}.
\end{equation}
Then, by replacing Eq.(\ref{Nphip}) into \eqref{v2}, it yields 
the following potential for any value of the parameter $p$
\begin{equation}
    V(\phi)=\frac{(p-1) (\phi-C_1)^{\frac{4(p-1)}{(4-p)}}}{A\,\gamma (p)^{-(p-1)}+B (p-1) (\phi-C_1)^{\frac{4(p-1)}{(4-p)}}}.\label{Vphip}
\end{equation}
In order to find concrete expressions for $V(\phi)$, we will 
consider the cases $p=3/2$ and $p=3$. At first, for $p=3/2$ 
Eq.(\ref{Vphip}) becomes
\begin{equation}
 V(\phi)=\dfrac{(\phi-C_1)^{4/5}}{\left(2^{17}/5^{4}\right)^{1/5}(A^6 \alpha^3)^{1/5}+B(\phi-C_1)^{4/5}}.\label{Vp1}
\end{equation}
As asymptotic limits of Eq.(\ref{Vp1}), first we observe that if $\left(2^{17}/5^{4}\right)^{1/5}(A^6\alpha^3)^{1/5}\gg B(\phi-C_1)^{4/5}$, $V(\phi)$ behaves as a power law potential
\begin{equation}
 V(\phi)\simeq \dfrac{(\phi-C_1)^{4/5}}{\left(2^{17}/5^{4}\right)^{1/5}(A^6\alpha^3)^{1/5}},
\end{equation}

while if $\left(2^{17}/5^{4}\right)^{1/5}(A^6\alpha^3)^{1/5}\ll B(\phi-C_1)^{4/5}$, it becomes a constant potential
\begin{equation}
    V\simeq\dfrac{1}{B}.
\end{equation}

Now, for $p=3$, the potential (\ref{Vphip}) results \begin{equation}
 V(\phi)=\dfrac{(\phi-C_1)^8}{2^{15}(A\alpha^2)^3+B(\phi-C_1)^8}.
\end{equation}
It can be seen that when $2^{15}(A\alpha^2)^3\gg B(\phi-C_1)^8$, the potential behaves as
\begin{equation}
   V(\phi)\simeq \frac{\left(\phi-C_1\right)^8}{2^{15}(A\alpha^2)^3}.\label{Vlarge}
\end{equation}
On the contrary, if $2^{15}(A\alpha^2)^3\ll B(\phi-C_1)^8$, the potential presents a constant behaviour
\begin{equation}
    V(\phi)\simeq \frac{1}{B}.\label{Vcte}
\end{equation}
Thus the effective potential given by Eqs.(\ref{Vlarge}) and (\ref{Vcte}) correspond
to the limit potential for small and large $\phi$, respectively.

\subsection{Cosmological perturbations}
Upon replacement of Eq.\eqref{AN} into the expression for the scalar power spectrum \eqref{Ps}, we 
find that the constraint on the integration constant $A$ for any value of $p$ and $n$ is given by
\begin{equation}
    A=\dfrac{\sqrt{2n-1}N^p}{12\pi^2M_{pl}^4\mathcal{P}_s}.\label{ApN}
\end{equation}
However, as we will see, the tensor-to-scalar ratio becomes different and we will consider the case $p=1$ and $p\neq 1$ separately for the situation in which $n=2$.

Regarding the case (a) $p=1$ and $n=2$ by using Eqs.\eqref{epsilonn}, the potential \eqref{v2b} and Eq.\eqref{rN} along with the constraint on $A$ given by (\ref{ApN}), we get
that the tensor-to-scalar ratio becomes
\begin{equation}
    r=\dfrac{8}{12B M_{pl}^4\,\pi^2 \,\mathcal{P}_s - \sqrt{3}\, N \ln N}.\label{rp1}
\end{equation}
On the other hand, for the case (b) $p\neq 1$ and $n=2$, but using the potential \eqref{v2} instead, the tensor-to-scalar ratio is expressed as follows
\begin{equation}
    r=\dfrac{8(p-1)}{12 B(p-1)M_{pl}^4\,\pi^2 \,\mathcal{P}_s + \sqrt{3} \,N}.\label{rp}
\end{equation}

The allowed values of the integration constant $A$ are found by evaluating Eq.(\ref{ApN}) by using $\mathcal{P}_s\simeq2.2\times 10^{-9}$ for a range of the number of $e$-folds in the range $50\leq N \leq 70$, while the upper bound on the tensor-to-scalar ratio from BICEP2/Keck array (BK14) data sets the lower bound on $B$. Accordingly, for the case (a) $p=1$ and $n=2$, from Eq.(\ref{rp1}) it is found the following lower limit on $B$
\begin{equation}
    B>\frac{8-\sqrt{3}r_0 N \ln N}{12\pi^2M_{pl}^4\mathcal{P}_s r_0},\label{Bp1}
\end{equation}
where $r_0$ denotes the current upper limit on the tensor-to-scalar ratio. Besides, from Eq.(\ref{rp}), the lower limit of
$B$ for the case (b) $p\neq 1$ and $n=2$ yields
\begin{equation}
  B>  \frac{8(p-1)-\sqrt{3} r_0 N}{12(p-1)M_{pl}^4\,\pi^2 \,\mathcal{P}_s r_0}.\label{Bp}
\end{equation}
As before, in order to obtain concretes values, we consider the cases $p=3/2$ and $p=3$. 

The resulting allowed values for the integration constant $A$ and the corresponding lower bounds on $B$ for the cases $p=1$, $p=3/2$ and $p=3$ in the range $50\leq N \leq 70$ are shown in Table \ref{tab:AB2}.
\begin{table}[H]
    \centering
    \captionsetup[subtable]{position = below}
    \captionsetup[table]{position=bottom}
    \begin{subtable}{0.48\textwidth}
        \begin{tabular}{|c|c|c|}
            \hline
            \hspace{5mm}$N$\hspace{5mm} & \hspace{5mm}$A\,[M_{pl}^{-4}]$\hspace{5mm} & \hspace{5mm}$B\,[M_{pl}^{-4}]$\hspace{5mm}  \\\hline
            50 & $3.32\times10^{8}$ &\, $B>2.15\times 10^9$\, \\
            60 & $3.99\times10^{8}$ & \,$B>2.49\times 10^9$\, \\
            70 & $4.65\times10^{8}$ & \,$B>2.83\times 10^9$ \,\\\hline
        \end{tabular}
        \caption{$p=1,n=2$}
    \end{subtable}%
    \begin{subtable}{0.48\textwidth}
       \centering
        \begin{tabular}{|c|c|c|}
            \hline
            \hspace{5mm}$N$\hspace{5mm} & \hspace{5mm}$A\,[M_{pl}^{-4}]$\hspace{5mm} & \hspace{5mm}$B\,[M_{pl}^{-4}]$\hspace{5mm}  \\\hline
            50 & $2.35\times10^{9}$ & \,$B>1.88\times 10^8$\, \\
            60 & $3.09\times10^{9}$ & \,$B>5.52\times 10^7$\, \\
            70 & $3.89\times10^{9}$ &\, $B>-7.78\times 10^7$\, \\\hline
        \end{tabular}
        \caption{$p=3/2,n=2$}
        \label{tab:dimGMM}
    \end{subtable}%
    \newline
    \begin{subtable}{0.48\textwidth}
        \begin{tabular}{|c|c|c|}
            \hline
            \hspace{5mm}$N$\hspace{5mm} & \hspace{5mm}$A\,[M_{pl}^{-4}]$\hspace{5mm} & \hspace{5mm}$B\,[M_{pl}^{-4}]$\hspace{5mm}  \\\hline
            50 & $8.31\times10^{11}$ &\, $B>6.87\times 10^8$\, \\
            60 & $1.44\times10^{12}$ & \,$B>6.53\times 10^8$\, \\
            70 & $2.28\times10^{12}$ & \,$B>6.20\times 10^8$ \,\\\hline
        \end{tabular}
        \caption{$p=3,n=2$}
    \end{subtable}%
    
    \caption{Results for the constraints on the integration constants $A$ and $B$ for the second attractor $n_s-1=-\frac{p}{N}$, using $p=1$ (top left panel), $p=3/2$ (top right panel) and $p=3$ (bottom panel) each for $n=2$ in k-inflation. Here we have used $\mathcal{P}_s\simeq2.2\times 10^{-9}$ and $r_0=0.036$.}\label{tab:AB2}
\end{table}

\begin{figure}[H]

 \centering
       \includegraphics[width=0.9\linewidth]{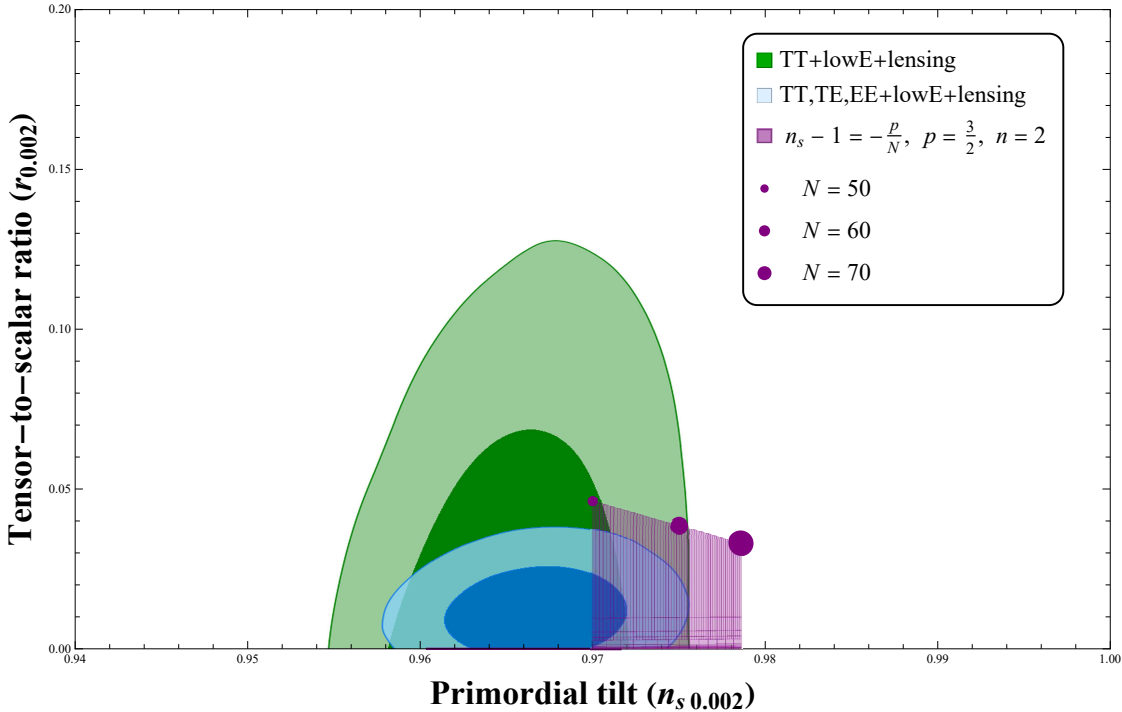}

    \caption{Plots of the tensor-to-scalar ratio $r$ against the scalar spectral index $n_s$ for the second attractor $n_s-1=-\frac{p}{N}$ along with the two-marginalized joint confidence contours for ($n_s,r$) at 68\% and 95\% C.L. from PLANCK 2018. Here we have considered the specific case in which $p=3/2$ for $n=2$.}\label{rns2}
\end{figure}

Following the same procedure of Section \ref{attractor1}, for the case (a) $p=1$ and $n=2$, the predictions in the $n_s-r$ plane may be obtained by plotting the attractor $n_s-1=-\frac{1}{N}$ and Eq.\eqref{rp1}, by varying simultaneously the integration constant $B$ in a wide range and 
the number of $e$-folds within the range $N=50-70$. However, we observe that the case $p=1$ yields values for the scalar spectral index (for $50\leq N \leq 70$) which are greater than its likelihood value. On the other hand, for the case (b) $p\neq1$ and $n=2$, we plot $n_s-1=-\frac{p}{N}$ and Eq.\eqref{rp}. In relation to the predictions for $p\geq 2$, the scalar spectral index now becomes smaller than its preferred value from current observations. In order to show concrete results, we will restrict ourselves to the case $p=3/2$. Fig.\ref{rns2} shows the $n_s-r$ plots for $p=3/2$ when $n=2$ considering the two-marginalized joint confidence at 68\% and 95\% C.L., from Planck 2018 results. We observe that for the case $p=3/2$, the purple shaded region lies within the 68\% and 95\% C.L. from BICEP/Keck for $50\leq N \leq 60$. These results show that for the power $n=2$ associated to the kinetic term, the attractor $n_s-1=-\frac{p}{N}$ only works for $p=3/2$.

\subsection{Reheating}

\begin{figure}[H]
    \includegraphics[scale=0.32]{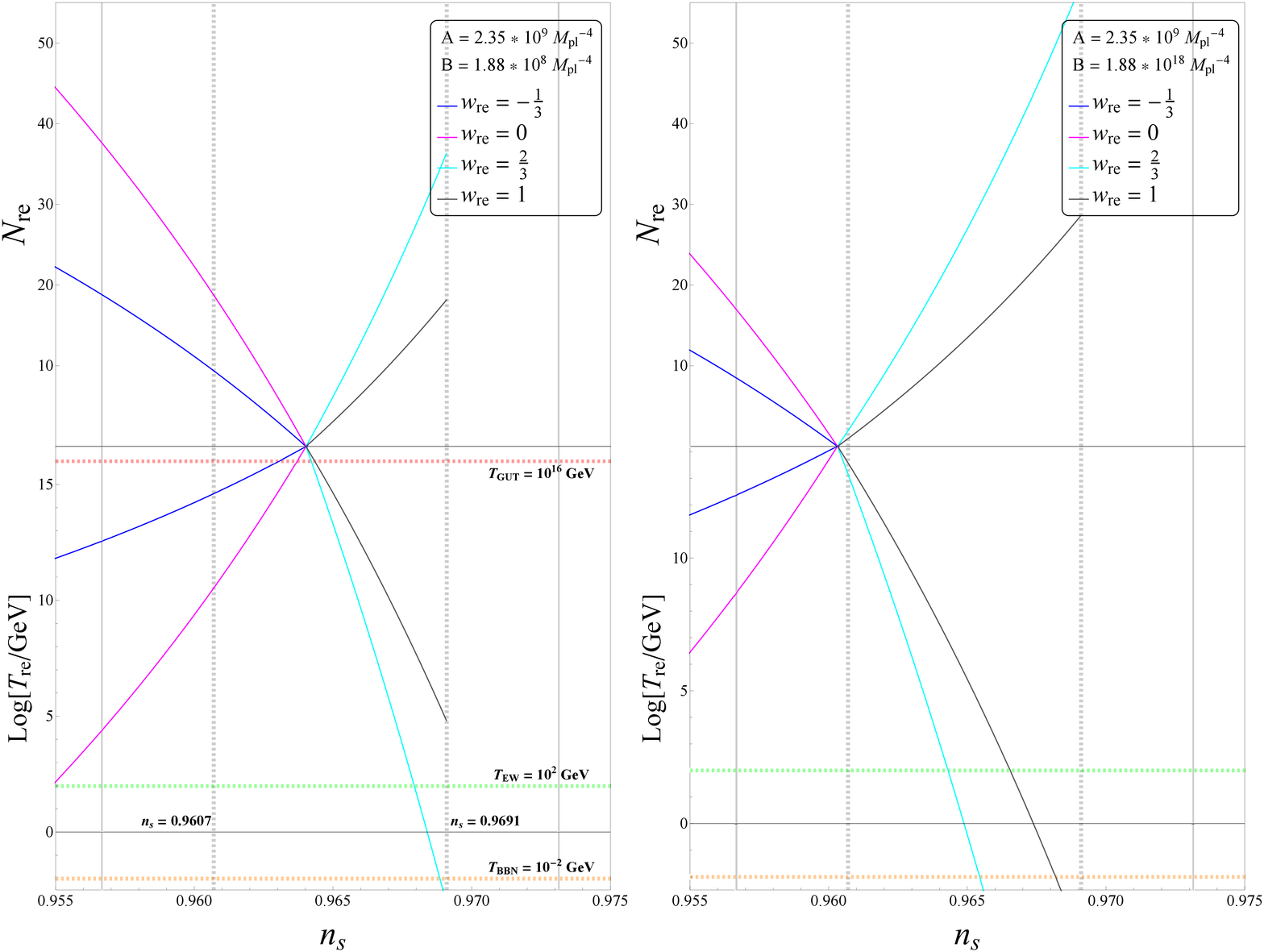}
    \caption{Plots of the reheating duration, $N_{re}$, and the reheating temperature $T_{re}$ for the power $n=2$ against the spectral index $n_s$ for the attractor $n_s-1=\frac{p}{N}$ and considering different values of the EoS parameter $w_{re}$. The blue, pink, cyan and black curves correspond to the EoS parameter $w_{re}=-1/3,0,2/3,1$, respectively. The gray dotted vertical lines indicates the observational constraints on the spectral index, i.e. $n_s=0.965\pm 0.004$. The green dotted horizontal line shows reheating temperatures being below the electroweak scale $T_{re}=10^{2}$ GeV, while the orange dotted line indicates the lower bound on $T_{re}$ from BBN, $T_{re}= 10^{-2}$ GeV.  For the upper and lower left panels we have used $A=2.35\times10^{9}\,M_{pl}^{-4}$ and $B=1.88\times10^{8}\,M_{pl}^{-4}$, while for the upper and lower right panels we have used $A=2.35\times10^{9}\,M_{pl}^{-4}$ and $B=1.88\times10^{18}\,M_{pl}^{-4}$.}
    \label{pntre3}
\end{figure}

Since the current CMB constraints only favor the case $p=3/2$, we proceed further to study the predictions of reheating phase regarding its duration through the number of $e$-folds $N_{re}$ as well as the temperature $T_{re}$. In doing so, we follow the same procedure as in Section \ref{attractor1}. Then, by plotting parametrically Eqs. \eqref{nre} and \eqref{tre} for the attractor $n_s-1=-\,\frac{3}{2N}$  for several values of the EoS parameter $w_{re}$ over the range $-\frac{1}{3}\leq w_{re}\leq 1$, we find the plots of Fig.\ref{pntre3}. Here, there are shown the plots for $N_{re}$ (upper plots) and $T_{re}$ (lower plots) against $n_s$. For the upper and lower left plots we have used the allowed value of $A$, given by $A=2.35\times10^{9}\,M_{pl}^{-4}$, and the minimum value of $B$ at $N=50$, i.e., $B=1.88\times 10^8\,M_{pl}^{-4}$ (see Table \ref{tab:AB2}). 

On the other side, for the upper and lower right plots we have used the same value of $A$ and an intermediate value of $B$, $B=1.88\times 10^{18}\,M_{pl}^{-4}$ at $N=50$. Firstly, from the $T_{re}$ against $n_s$ plots, it is found that as the value of $B$ increases, the maximum reheating temperature decreases. Secondly, by performing the same analysis for the EoS parameter as the previous section, we found that for all the four values of $w_{re}$ it is required that the number of $e$-folds be $N<50$ in order to achieve a reheating temperature above the electroweak scale and below the GUT scale. As a consequence, the scalar spectral index must be lower than $0.967$, which is not consistent with the predictions of the model on the $n_s-r$ plane. Accordingly, in despite the attractor $n_s-1=-\,p/N$ becomes supported by current CMB data only for $p=3/2$, this is not consistent with reheating constraints.


\section{Third attractor $n_s-1=-\dfrac{\beta}{N^q}$}\label{attractor3}
\subsection{Dynamics of inflation}

Here, we introduce a more general attractor, namely \cite{Huang:2007qz}
\begin{equation}
n_s-1=-\dfrac{\beta}{N^q}, \label{nsq}
\end{equation}
which is a generalization of the previous expression $n_s-1=-\,p/N$. Here, $\beta$ and $q$ are constant parameters. From Eq.\eqref{ns}, we found
\begin{equation}
    -\dfrac{\beta}{N^q}=\left(\ln\dfrac{V_{,N}}{V^2}\right)_{,N}.
\end{equation}
A first integration of this equation with respect to the number of $e$-folds $N$ yields
\begin{equation}
    \dfrac{V_{,N}}{V^2}=Ae^{-\beta N^{1-q}/(1-q)}\label{vn3},
\end{equation}
where $A$ is a positive integration constant, since we have assumed that $V_{,N}>0$. Integrating once again with respect to $N$, we found the following expression for the potential $V(N)$
\begin{equation}
    V(N)=\dfrac{1-q}{A\,N\left(\frac{1-q}{\beta\,N^{1-q}}\right)^{\frac{1}{1-q}}\Gamma\left(\frac{1}{1-q},\frac{\beta\,N^{1-q}}{1-q}\right) + B(1-q)},\,\,\textup{with}\,q\neq 1.\label{v3}
\end{equation}
where $B$ is an arbitrary integration constant and $\Gamma\left(\frac{1}{1-q},\frac{\beta\,N^{1-q}}{1-q}\right)$ represents the incomplete gamma function \cite{gamma}. The expression for the scalar field $\phi$ in terms of the
the number of $e$-folds $N$ is obtained by using Eq.\eqref{phiN} and the potential \eqref{v3}, yielding
\begin{equation}
    \phi(N)-C_1=\int \alpha^{\frac{2n-1}{2n}}\left(Ae^{-\frac{\beta N^{1-q}}{1-q}}\left(\dfrac{1-q}{A\,N\left(\frac{1-q}{\beta\,N^{1-q}}\right)^{\frac{1}{1-q}}\Gamma\left(\frac{1}{1-q},\frac{\beta\,N^{1-q}}{1-q}\right) + B(1-q)}\right)^{2-n}\right)^{\frac{1}{2n}}\, dN,\label{dphi3}
\end{equation}
where $C_1$ is a new integration constant. It must be noted that the integral on the right hand side of Eq.\eqref{dphi3} only has analytical solutions for two cases: (a) $q=0$ and $n=1$; (b) $q=0$ and $n=2$. The case (a) corresponds to a constant attractor in standard canonical inflation, which has been studied so far \cite{chiba}, so we will restrict ourselves to the case (b).  For this latter case, which also corresponds to a constant attractor, Eq.(\ref{v3}) becomes
\begin{equation}
    V(N)=\dfrac{\beta}{Ae^{-\beta N}+\beta B}\label{v3q0}.
    \end{equation}
On the other hand, for the case b) i.e., $q=0$ and $n=2$, Eq.(\ref{dphi3}) can be inverted in order to find the number $e$-folds as a function
of the the scalar field as follows
\begin{equation}
    N(\phi)=-\dfrac{4}{\beta}\ln\left(-\dfrac{\beta (\phi-C_1)}{4(\alpha^3 A)^{1/4}}\right).
\end{equation}
Accordingly, if we replace this expression into Eq.\eqref{v3q0}, the potential as a function of the scalar field yields
\begin{equation}\label{v3q0n2}
    V(\phi)=\dfrac{\beta}{\left(-\frac{\beta}{4}\right)^4 \frac{(\phi-C_1)^4}{\alpha^3}+\beta B}.
\end{equation}
In particular, if  $\left(\frac{-\beta}{4}\right)^4 \frac{(\phi-C_1)^4}{\alpha^3\beta B}\ll 1$ the potential \eqref{v3q0n2} becomes
\begin{equation}
    V(\phi)\simeq\dfrac{1}{B}\left[1-\left(-\frac{\beta}{4}\right)^4 \frac{(\phi-C_1)^4}{\alpha^3\beta B}\right],
\end{equation}
which has the form of a quartic Hilltop inflationary model \cite{Boubekeur:2005zm} as long as $B>0$.

Since we don't have analytical expressions for the number of $e$-folds as a function of the scalar field, then the effective potential $V(\phi)$ can't be obtained for values $q$ such that $q\neq 0$. However, we can find the effective potential in terms of $N$. Thus, we can write down the potential as a function of the number of $e$-folds for $q=2$ and $q=3$ when $n=2$, which become
\begin{eqnarray}
  V(N)&=&\dfrac{1}{A\left(\beta\, \textup{Ei}\left(\frac{\beta}{N}\right)-e^{\frac{\beta}{N}}\right)-B},
  \label{v3q2}\\
   V(N)&=&\dfrac{2}{A\left(\sqrt{2\pi\,\beta}\,\textup{erfi}\left(\dfrac{\sqrt{\beta}}{\sqrt{2}\,N}\right)-2e^{\frac{\beta}{2N^2}}\,N\right)-2B},\label{v3q3}
\end{eqnarray}
 respectively. In Eq.(\ref{v3q2}), $\textup{Ei}(x)$ represents the exponential integral function defined as \cite{Ei}
\begin{equation}
    \textup{Ei}(x)\equiv -\int _{-x}^{\infty}\frac{e^{-y}}{y} dy.
\end{equation}
 Additionally, $\textup{erfi}(x)$ in Eq.(\ref{v3q3}) corresponds to the imaginary error function \cite{erfi}. In this context, we can't reconstruct the effective potential as a function of the scalar field. However, for each of these expressions along with Eq.\eqref{v3q0}, the power spectrum as well as the tensor-to-ratio can be computed in order to constraint the parameters by using the CMB constraints on the inflationary observables.

\subsection{Cosmological perturbations}

By replacing Eq.\eqref{vn3} into the expression for the scalar power spectrum \eqref{Ps} and solving for the integration constant $A$, it is  found that
\begin{equation}
    A=\dfrac{e^{-\beta N^{1-q}/1-q}}{12\pi^2M_{pl}^4c_s\mathcal{P}_s}.\label{Aq}
\end{equation}
Furthermore, the several expressions for the tensor-to-scalar ratio for the potentials $V(N)$, namely 
\eqref{v3q0}, \eqref{v3q2} and \eqref{v3q3}, are found by replacing  the former equations and \eqref{epsilonn} into Eq. \eqref{rN}. Thus, by using the constraint on the integration constant $A$ given by
Eq.(\ref{Aq}), it yields
\begin{eqnarray}
   r&=&\dfrac{8\,c_s\beta}{1+12B\,e^{2N\beta}\pi^2M_{pl}^4c_s\mathcal{P}_s},\label{rq0}\\
   r&=&\dfrac{8\,c_s e^{\beta N}}{12B\pi^2M_{pl}^4c_s\mathcal{P}_s -e^{\frac{\beta}{N}}N+\beta\textup{Ei}\left(\frac{\beta}{N}\right)},\label{rq2}\\
  r&=&\dfrac{16\,c_s} {24B\pi^2M_{pl}^4c_s\mathcal{P}_s-2N+\sqrt{2\pi \beta}e^{-\frac{\beta}{2N^2}}\textup{erfi}\left(\frac{\sqrt{\beta}}{\sqrt{2}N}\right)},\label{rq3}
\end{eqnarray}
for the specific cases $q=0$, $q=2$ and $q=3$, respectively. For $q\geq 4$ it is found that a negative value of $\beta$ is needed for $r$ to be positive, which implies $n_s>1$ and the attractor does not work.

%
%
In a similar way as for the attractor (i), the CMB constraints on the inflationary observables are used to find
the allowed values for the parameter $\beta$ appearing in the expression for this last expression for $n_s(N)$, and also for the integration constants $A$ and $B$. At first, the value of $\beta$ is constrained so that over the range of the number of $e$-folds $50\leq N\leq 70$, the scalar spectral index $n_s$ becomes consistent with current observational bounds $n_s=0.965\pm 0.004$. In this regard, the allowed values for $\beta$ are within the ranges $65\leq\beta\leq 190$ and $3.2\times 10^2\leq\beta\leq 9.6 \times 10^2$ for $q=2$ and $q=3$, respectively. For the special case $q=0$, i.e. the constant attractor, it is found that $\beta=0.0351$. Secondly, we use
the constraint on the amplitude of the scalar spectrum $\mathcal{P}_s\simeq2.2\times 10^{-9}$ to set $A$ in (\ref{Aq}), while the allowed values of $B$ are found from the upper bound on $r$ by the BICEP/Keck data, i.e. $0<r<0.036$ \cite{BICEP:2021xfz}. In doing so, if $r_0$ denotes the upper limit on the tensor-to-scalar ratio, from, Eqs.(\ref{rq0}), (\ref{rq2}), and (\ref{rq3}), we found the following lower bounds on $B$
\begin{eqnarray}
    B&>& \frac{8\,c_s \beta-r_0}{12 e^{2N\beta}\pi^2M_{pl}^4c_s\mathcal{P}_s r_0},\label{Bq0}\\
    B&>& \frac{8\, c_s e^{\beta N}+e^{\frac{\beta}{N}}N r_0-\beta\,r_0 \textup{Ei}\left(\frac{\beta}{N}\right) }{12 \pi^2M_{pl}^4c_s\mathcal{P}_s r_0},\label{Bq2}\\
    B&>& \frac{16\,c_s+2N r_0-\sqrt{2\pi \beta} e^{-\frac{\beta}{2 N^2}} \textup{erfi}\left(\frac{\sqrt{\beta}}{\sqrt{2}N}\right)}{24 \pi^2M_{pl}^4c_s\mathcal{P}_s r_0},\label{Bq3}
\end{eqnarray}
for $q=0$, $q=2$ and $q=3$, respectively. Table \ref{tab:AB3} summarizes the allowed values for the integration constants $A$ and $B$ for $n=2$ and several values of $q$ by considering the CMB constraints on the inflationary observables.

\begin{table}[H]
    \centering
    \captionsetup[subtable]{position = below}
    \captionsetup[table]{position=bottom}
    \begin{subtable}{0.48\textwidth}
        \begin{tabular}{|c|c|c|}
            \hline
            \hspace{5mm}$N$\hspace{5mm} & \hspace{5mm}$A\,[M_{pl}^{-4}]$\hspace{5mm} & \hspace{5mm}$B\,[M_{pl}^{-4}]$\hspace{5mm}  \\\hline
            50 & $3.84\times10^{7}$ & \,$B>6.63\times 10^8$\, \\
            60 & $5.46\times10^{7}$ & \,$B>6.63\times 10^8$\, \\
            70 & $7.76\times10^{7}$ & \,$B>6.63\times 10^8$\,\\\hline
        \end{tabular}
        \caption{$q=0,\,n=2, \beta=0.0351$}
    \end{subtable}%
    \hspace{3mm}
    \begin{subtable}{0.48\textwidth}
        \begin{tabular}{|c|c|c|}
            \hline
            \hspace{5mm}$N$\hspace{5mm} & \hspace{5mm}$A\,[M_{pl}^{-4}]$\hspace{5mm} & \hspace{5mm}$B\,[M_{pl}^{-4}]$\hspace{5mm}  \\\hline
            50 & $7.37\times10^{5}$ & \,$B>7.21\times 10^8$\, \\
            57 & $9.65\times10^{5}$ & \,$B>7.33\times 10^8$\, \\
            65 & $1.22\times10^{6}$ & \,$B>7.60\times 10^8$\,\\\hline
        \end{tabular}
        \caption{$q=2,\,n=2, \beta=110$}
    \end{subtable}%
    \newline
    \begin{subtable}{0.48\textwidth}
        \begin{tabular}{|c|c|c|}
            \hline
            \hspace{5mm}$N$\hspace{5mm} & \hspace{5mm}$A\,[M_{pl}^{-4}]$\hspace{5mm} & \hspace{5mm}$B\,[M_{pl}^{-4}]$\hspace{5mm}  \\\hline
            50 & $2.21\times10^{6}$ & \,$B>8.13\times 10^8$\, \\
            55 & $2.68\times10^{6}$ & \,$B>8.42\times 10^8$\, \\
            65 & $3.10\times10^{6}$ & \,$B>8.75\times 10^8$\,\\\hline
        \end{tabular}
        \caption{$q=3,\,n=2, \beta=5.5 \times 10^2$}
    \end{subtable}%
    \caption{Results for the constraints on the integration constants $A$ and $B$ for the third attractor $n_s-1=-\frac{\beta}{N^q}$ in k-inflation.}\label{tab:AB3}
\end{table}

\begin{figure}[H]\setcounter{subfigure}{0}
\centering
    \begin{subfigure}{\textwidth}
    \centering
        \includegraphics[width=0.9\linewidth]{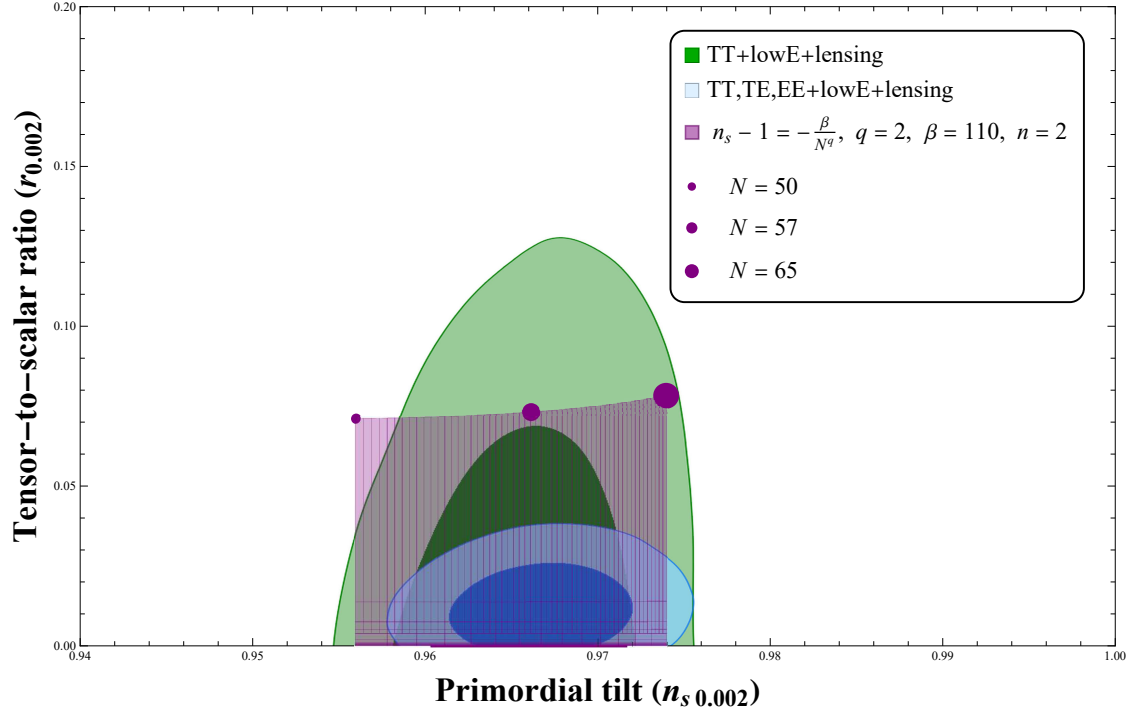}
        \caption{$q=2,\,\beta=110,\,n=2$}
    \end{subfigure}%
    \\
    \begin{subfigure}{\textwidth}
    \centering
        \includegraphics[width=0.9\linewidth]{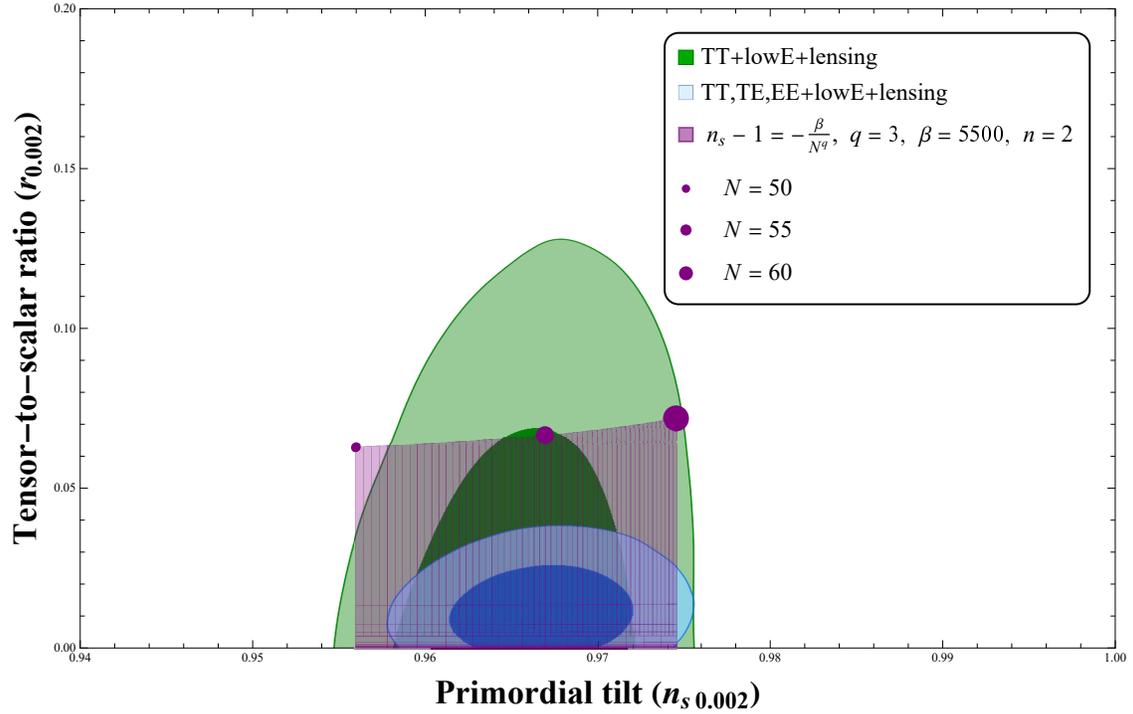}
        \caption{$q=3,\,\beta=5500,\,n=2$}
    \end{subfigure}%
    \caption{Plots of the tensor-to-scalar ratio $r$ against the scalar spectral index $n_s$ for the second attractor $n_s-1=-\frac{\beta}{N^q}$ along with the two-marginalized joint confidence contours for ($n_s,r$) at 68\% and 95\% C.L. from BICEP3.}\label{rns3}
\end{figure}

In the same way as in Section \ref{attractor1}, we generate the predictions on the $n_s-r$ plane by plotting the attractor $n_s-1=-\frac{\beta}{N^q}$ and the respective equations of $r$ (Eqs. \eqref{rq2} and \eqref{rq3}) by varying simultaneously the integration constant $B$ in a wide range and $N$ within the range $N=50-70$. Fig. \eqref{rns3} shows the plots for $q=2$ and $q=3$ both for $n=2$ considering the two-marginalized joint confidence at 68\% and 95\% C.L., from BICEP/Keck data. It should be noted that for the constant attractor $q=0$, the curve on the $n_s-r$ plane becomes a vertical line at the central value of $n_s$ (not shown).



Regarding the reheating predictions for this last attractor, the equation for the slow-roll parameter $\epsilon$ at the end of inflation, i.e., $\epsilon (\phi_{end})\equiv 1$ cannot be solved analytically in order to obtain $\phi_{end}$, and therefore there is no possible to find an analytical expression for $V_{end}$, which is needed to compute $N_{re}$ and $T_{re}$ as well. As a consequence, a numerical study of reheating constraints is needed instead, which is beyond the goals of the present work.

\section{Conclusions}\label{conclusions}

In this paper we have applied the reconstruction scheme from the attractor points $n_s(N)$ to the framework of k-inflation, in particular for a non-linear kinetic term $K(X)=k_{n+1}X^n$, in order to rebuild the potential associated to the inflaton field $V(\phi)$. In doing so, we have considered the following attractors: (i) $n_s-1=-\frac{2}{N}$, (ii) $n_s=1-\frac{p}{N}$, and (iii) $n_s-1=-\frac{\beta}{N^q}$. 
For simplest the attractor (i), it was possible to find analytical expressions for the reconstructed inflaton potential $V(\phi)$ for the non-canonical scalar field $n=2$. In particular, for $n=1$ we recovered the T-model corresponding to the standard canonical case. Regarding the case $n=2$, the resulting potential interpolates between a chaotic quadratic and a constant effective potential, and the allowed ranges for the parameters characterizing this model were obtained by comparing its predictions on the $n_s-r$ plane with current CMB constraints on the inflationary observables. As a further analysis, we studied the reheating constraints regarding the duration of this phase from $e$-folds $N_{re}$ and the temperature $T_{re}$ in order to narrow the parameter space of the reconstructed model. In doing so, we considered several values of the EoS parameter of the fluid into which the inflaton decays over a range $-\frac{1}{3}\leq w_{re}\leq 1$. In first place, by restricting the number of $e$-folds $N$, it was possible to find an allowed range for the reheating temperature $T_{re}$ for each parameter of state $w_{re}$. We also found that the instantaneous reheating temperature, in which $N_{re}\sim 0$, and it becomes $T_{re}\simeq 10^{16}$ GeV, which is similar to those found for standard inflation scenario \cite{paper1}. Once the number of $e$-folds was restricted for each EoS (Table \ref{wn1}), we plot the corresponding predictions in the $r-n_s$ plane, and it was found that all the  EoS parameters considered are preferred by current observational bounds. For
the attractor (ii), analytical solutions for $V(\phi)$ were found for $p=1$, $p=3/2$, and $p=3$ when $n=2$. Specifically, for $p=1$, the inflaton potential obtained is similar to those from loop
inflation (LI) model. On the other hand, for $p=3/2$ and $p=3$, the resulting potential can be either a power-law or constant. Regarding its predictions on the $n_s-r$ plane, this attractor is not supported by current observational constraints on the scalar spectral index $n_s$ for $p=1$ and $p=3$, since $n_s$ becomes greater and lower than its likelihood value, respectively. Nevertheless, for the case $p=3/2$, its predictions are within the 68\% and 95\% C.L. from BICEP/Keck for $50\leq N \leq 60$. Although a further analysis regarding the reheating  phase is possible for this attractor, the corresponding constraint
on the number of $e$-folds $N$ becomes small ($N<50$) for any value of the EoS parameter $w_{re}$, and then this second attractor for $p=3/2$ does not work. For the attractor (iii),
we were able to find analytical solutions for $V(\phi)$ 
only for the case $q=0$ and $n=2$, corresponding to a constant attractor. In that case, the resulting inflaton potential
can be approximated to a quartic hilltop one. For this constant attractor, the parameter $\beta$ can be 
restricted in order to reproduce the observational value for the scalar spectral index. For completeness, we also obtained $V(N)$ analytically for two additional cases, namely $q=2$ for $n=2$, and $q=3$ for $n=2$ in order to study their  predictions on the $n_s-r$ plane. Following the same procedure for the constant attractor, we also obtained the allowed range for $\beta$ for the cases $q=2$ and $q=3$. The viable values for the rest of parameters, i.e., the integration constants $A$ and $B$ for the three cases, they were found by using the CMB constraints on the scalar power spectrum and the tensor-to-scalar ratio, respectively (summarized in Table \ref{tab:AB3}). Concerning the reheating predictions for this last attractor, we haven't been able to find analytical results, then a numerical analysis is needed instead of an analytical one, which goes beyond the main goals of this work. 

As a final remark, we have not addressed the reconstruction scheme from the tensor-to-scalar ratio $r(N)$ or the slow-roll parameter $\epsilon(N)$, and neither a more general expression for the Lagrangian density, e.g. $P(\phi,X)=\sum_{n\geq 0} g_n(\phi) X^{n+1}-V(\phi)$ in our analysis. We hope to be able to address these points in a future work.

\section*{Acknowledgements}

C.O. is supported by the Pontificia Universidad Católica de Valparaíso trough the scholarship ``Beca Término de Tesis''.

\end{document}